%% file: main.tex
\documentclass[sigconf]{acmart}
\AtBeginDocument{%
  }

\copyrightyear{2026}
\acmYear{2026}
\setcopyright{cc}
\setcctype{by}
\acmConference[CHI '26]{Proceedings of the 2026 CHI Conference on Human Factors in Computing Systems}{April 13--17, 2026}{Barcelona, Spain}
\acmBooktitle{Proceedings of the 2026 CHI Conference on Human Factors in Computing Systems (CHI '26), April 13--17, 2026, Barcelona, Spain}
\acmPrice{}
\acmDOI{10.1145/3772318.3790634}
\acmISBN{979-8-4007-2278-3/2026/04}
\usepackage{xcolor}
\usepackage{enumitem}

\usepackage{amsmath}
\usepackage{subcaption}

\DeclareMathOperator*{\argmin}{arg\,min}

\newsavebox{\promptboxbox}
\newenvironment{promptbox}[1]{%
  \begin{lrbox}{\promptboxbox}%
  \begin{minipage}{0.97\linewidth}%
  \textbf{#1}\par\medskip
}{%
  \end{minipage}%
  \end{lrbox}%
  \begin{center}%
  \setlength{\fboxsep}{6pt}%
  \setlength{\fboxrule}{0.8pt}%
  \fcolorbox{gray!75!black}{gray!5!white}{\usebox{\promptboxbox}}%
  \end{center}%
}

\begin{document}

\title{ViSTAR: Virtual Skill Training with Augmented Reality with 3D Avatars and LLM coaching agent}



\author{Chunggi Lee* }
\email{chunggi\_lee@g.harvard.edu}
\orcid{0000-0002-6164-2563}
\affiliation{%
  % \institution{Visual Computing Group}
  \institution{Harvard University}
  \city{Cambridge}
  \state{Massachusetts}
  \country{USA}
}

\author{Hayato Saiki*}
\email{saiki@ai.iit.tsukuba.ac.jp}
\orcid{0009-0005-8374-9483}
\affiliation{%
  % \institution{Artificial Intelligence Laboratory}
  \institution{University of Tsukuba}
  \city{Tsukuba}
  \country{Japan}
}

\author{Tica Lin}
\email{mlin@g.harvard.edu}
\orcid{0000-0002-2860-0871}
\affiliation{%
  \institution{Dolby Laboratories}
  \city{Atlanta}
  \state{Georgia}
  \country{USA}
}

\author{Eiji Ikeda}
\email{ikeda.eiji.ga@u.tsukuba.ac.jp}
\orcid{0000-0003-0420-4564}
\affiliation{%
  \institution{}
  \institution{University of Tsukuba}
  \city{Tsukuba}
  \country{Japan}
}

% \author{Kaori Tachibana}
% \email{tachibana@ipu.ac.jp}
% \orcid{0000-0002-1669-0689}
% \affiliation{%
%   \institution{Ibaraki Prefectural University of Health Sciences}
%   \city{Ibaraki}
%   \country{Japan}
% }

\author{Kenji Suzuki}
\email{kenji@ieee.org}
\orcid{0000-0003-1736-5404}
\affiliation{%
  % \institution{Institute of Systems and Information Engineering}
  \institution{University of Tsukuba}
  \city{Tsukuba}
  \country{Japan}
}

\author{Chen Zhu-Tian}
\email{ztchen@umn.edu}
\orcid{0000-0002-2313-0612}
\affiliation{%
  \institution{University of Minnesota-Twin Cities}
  \city{Minneapolis, Minnesota}
  \country{USA}
}

\author{Hanspeter Pfister}
\email{pfister@seas.harvard.edu}
\orcid{0000-0002-3620-2582}
\affiliation{%
  % \institution{Visual Computing Group}
  \institution{Harvard University}
  \city{Cambridge}
  \state{Massachusetts}
  \country{USA}
}

\begin{abstract}
We present \textbf{\name}, a \textbf{Vi}rtual \textbf{S}kill \textbf{T}raining system in \textbf{AR} that supports self-guided basketball skill practice, with feedback on balance, posture, and timing. From a formative study with basketball players and coaches, the system addresses three challenges: understanding skills, identifying errors, and correcting mistakes. \namespace follows the Behavioral Skills Training (BST) framework—instruction, modeling, rehearsal, and feedback. It provides feedback through visual overlays, rhythm and timing cues, and an AI-powered coaching agent using 3D motion reconstruction. We generate verbal feedback by analyzing spatio-temporal joint data and mapping features to natural-language coaching cues via a Large Language Model (LLM). A key novelty is this feedback generation: motion features become concise coaching insights. In two studies (N=16), participants generally preferred our AI-generated feedback to coach feedback and reported that \namespace helped them notice posture and balance issues and refine movements beyond self-observation.

% In two exploratory studies (N=16), participants generally preferred our AI-generated feedback to coach feedback and reported that \namespace helped them notice posture and balance issues and consider how to adjust their movements beyond self-observation.
\end{abstract}




\begin{teaserfigure}
  \includegraphics[width=\textwidth]{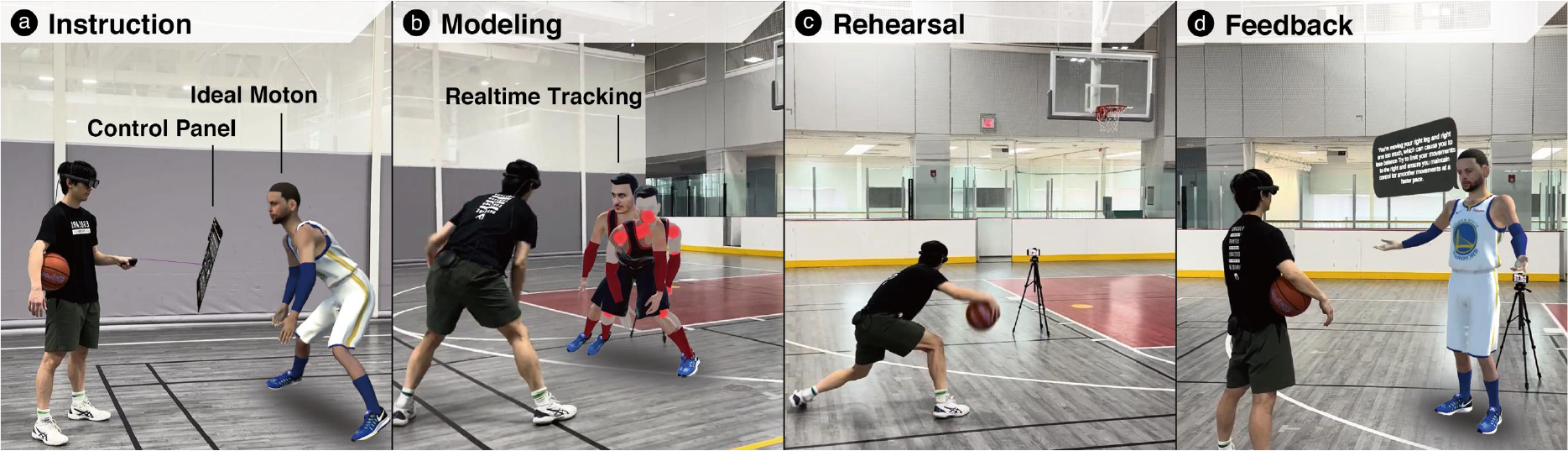}
  \caption{ \namespace is designed based on the Behavioral Skills Training (BST) teaching framework with four key steps: instruction, modeling, rehearsal, and feedback. Learners can (a) receive instruction of ideal motion demonstrated by a 3D animated avatar, (b) break down the action into key segments with real-time expert overlays, (c) rehearse the movement by recording their own performance, and (d) receive comprehensive multi-faceted feedback from a virtual coach in verbal and visual form.}
  \Description{Figure 1: Overview of the ViSTAR system design based on the Behavioral Skills Training framework. The figure illustrates four stages: instruction using a 3D animated avatar demonstrating ideal motion, modeling with segmented expert overlays, rehearsal through user-recorded performance, and feedback provided by a virtual coach using visual and verbal cues.
}
  \label{fig:teaser}
\end{teaserfigure}

\makeatletter
\def\blfootnote{\gdef\@thefnmark{}\@footnotetext}
\makeatother


\newcommand{\name}{ViSTAR}
\newcommand{\namespace}{ViSTAR }

\newcommand{\cglee}[1]{\textcolor{black}{{#1}}}
\newcommand{\cgleetwo}[1]{\textcolor{black}{{#1}}}

\definecolor{posture}{HTML}{424B9D}
\definecolor{verbal}{HTML}{44B986}
\definecolor{movement}{HTML}{C43B3A}
\definecolor{path}{HTML}{CA7C3C}
\maketitle
\blfootnote{*These authors contributed equally to this research.}

\input{Source/1_introduction}
\input{Source/2_related_work}

\input{Source/3_formative_study}
\input{Source/4_design_consideration}

\input{Source/5_system}

\input{Source/6_user_study}

\input{Source/7_discussion}

\input{Source/8_conclusion}

\begin{acks}
This work is supported by NSF grant CRCNS-2309041 and Harvard Data Science Initiative Trust in Science Fund Award.
\end{acks}

\bibliographystyle{ACM-Reference-Format}
\bibliography{references}

\appendix
\input{Source/9_appendix}

\end{document}

%% file: Source/1_introduction.tex
\section{INTRODUCTION}

\cglee{Motor skill acquisition is widely described as an embodied process across domains such as dance, music, craft, and sports \cite{wilson2002six, barsalou2008grounded, bailey2010body, wood2009tacit, leman2007embodied, juntunen2020ways}.  In this view, learners gradually align how movements \textit{feel} with how they are actually \textit{performed}, cultivating an internal sense of balance, rhythm, weight transfer, and timing. In competitive sports, these embodied skills directly shape athletes’ ability to create scoring opportunities and respond to rapidly changing game situations. Because these skills are grounded in internal sensations rather than explicit rules, athletes often find it difficult to identify and correct their own technique.  As a result, motor skill acquisition is considered highly relevant to coaching practice, with significant practical implications for training and performance\cite{fullagar2019practitioner, weng2025bridging}.}

Embodied motor learning is fundamental for both recreational and elite athletes \cite{verburgh2016key, krause2019enhancing, tribolet2022does}. While coach-guided training effectively translates internal sensations into actionable techniques \cite{hodges2002modelling, cote2007practice}, such personalized guidance is resource-intensive and largely inaccessible to non-professional athletes \cite{otte2019skill}. \cglee{Timely, precise feedback is critical for refining technique~\cite{lin2021towards, weng2025bridging, cheng2024viscourt, jheng2025badminton}, yet remains largely inaccessible for complex and multi-step movements. To support motor learning without a coach, prior research has leveraged head-mounted displays (HMDs)-based immersive training across various sports and dance~\cite{gong2024volleynaut, cheng2024viscourt, jheng2025badminton, chen2022vcoach, lin2021towards, esaki2024efficient}. While promising for tactical decision-making or repeatable skills (e.g., free-throws), current HMDs introduce practical constraints in high-intensity sports (e.g., fast movements and limited field of view), which limit their ecological feasibility in full gameplay. As a result, complex motor skills that involve multiple phases and coordinated limb actions (e.g., advanced basketball dribbling and footwork) remain relatively underexplored in Augmented Reality (AR)/Virtual Reality (VR) coaching systems.}

\cglee{
Motivated by the gap in providing guidance, we investigate an AR- and AI-driven system. This system focuses on kinematics, utilizing external feedback as a scaffold for the athlete's own reflection on their internal bodily awareness, thereby helping them recognize and interpret technical errors. We focus on controlled, demanding basketball moves where HMD use is practical, employing Behavioral Skills Training (BST)~\cite{sarokoff2004effects} as a pedagogical scaffold to sequence multi-phase AR guidance. Our design goal is to use these rich external cues (e.g., joint angles and posture) to help athletes reflect on their felt sense of balance, rhythm, and weight transfer. Critically, \namespace provides kinematic feedback and verbal guidance to scaffold this reflection, but we explicitly do not attempt to directly measure or train inner bodily sensations, which remain outside the reach of current non-invasive hardware.
}
Through a user-centered design process, we developed \name, an AR training system that provides basketball learners with personalized coaching feedback for isolated motor skills using 3D avatars and multi-faceted, LLM-powered feedback. At a high level, \namespace adopts the four components of BST—\textbf{\textit{instruction, modeling, rehearsal}}, and \textbf{\textit{feedback}}~\cite{sarokoff2004effects}—as the backbone of the training workflow. In the \textbf{\textit{instruction}} and \textbf{\textit{modeling}} phases, \name uses 3D avatars and segmented exemplars to present a 360° view of the move and support step-by-step visual alignment. During \textbf{\textit{rehearsal}}, learners practice the move while \namespace captures their motion, and in the \textbf{\textit{feedback}} phase, we provide side-by-side comparison, hit judgments, and verbal feedback that highlight which parts of the motion need adjustment. In designing feedback, we explore two levels of visual guidance: \textbf{(1) Holistic Motion}, which summarizes overall movement quality with a focus on flow and timing, and \textbf{(2) Localized Motion}, which targets specific motion segments (Sec.~\ref{sec:design}). Because verbal feedback is central to effective coaching, we also contribute an LLM-powered pipeline that translates joint-level motion analysis into natural-language coaching cues by (1) transforming motions into textual descriptors, (2) using Random Forest decision paths to identify salient motion differences, and (3) leveraging an LLM to generate actionable, context-aware feedback. To address the lack of datasets containing paired motions of the same skill, we simulate learner executions by injecting joint-level perturbations and use Random Forest to prioritize joint differences, making this data-to-language transformation feasible and scalable.

\cglee{We conducted two user studies to examine the quality of AI-generated coaching feedback and the perceived usefulness of the system. The first compared participants’ preferences between AI and human coach feedback, finding participants preferred AI for identifying posture errors and providing concrete, actionable corrections. The second study compared traditional self-observation with \name's AR system (integrating a 3D avatar and LLM guidance). While the small sample size limits strong claims about performance improvement, participants reported that \namespace helped them understand and recognize errors, rating the AR training as both engaging and useful.}

\cglee{In summary, we make the following contributions: (1) \name, an AR skill training system leveraging reconstructed 3D avatars and multi-faceted feedback for basketball practice; (2) a feedback design framework combining holistic (flow/rhythm) and localized (joint/path) guidance to visualize posture, timing, and weight transfer; (3) a method for generating AI coaching feedback by linking joint-level motion analysis with Large Language Models; and (4) an empirical evaluation (\name) through two user studies, providing insights into AI vs. coach feedback and how AR guidance supports self-directed learning. Given the small-sample, short-term constraints, we position \namespace as an enabling AR+AI platform that supports reflection on embodied aspects of skill learning, rather than a definitive performance-enhancing tool. }

%% file: Source/2_related_work.tex
\section{RELATED WORK}
%% target at 1 page

\subsection{Sports Training in Immersive Systems}
Immersive technologies such as Virtual Reality (VR) and Augmented Reality (AR) have been widely explored for motor skill training in sports. These systems are typically categorized into VR-based and AR-based approaches.

\textbf{VR-based training systems} have gained significant traction for their high immersion and control, with several studies showing their effectiveness in training tactical and perceptual skills in basketball.
For example, Tsai et al.~\cite{tsai2017train} developed a system that allows players to rehearse basketball tactics from global or player-specific perspectives, with virtual defenders reacting to head pose.  In a follow-up study, they~\cite{tsai2020feasibility} used a multi-camera setup and VR device to evaluate performance, finding that increased immersion enhances understanding of offensive strategies and strategic imagery.  
VisionCoach~\cite{liu2023visioncoach} trains visual scanning and passing decisions through VR-based game scenarios.  A number of review studies have thoroughly examined the use of VR in sports training, focusing on performance assessment and training in team ball sports~\cite{akbacs2019application}, applications for competitive athletes~\cite{faure2020virtual}, and the integration of interactive VR systems in sport~\cite{neumann2018systematic}. While promising in controlled environments, VR systems replace the physical training context, which can limit realistic movement, hinder transfer to actual courts, and raise safety concerns.  
These limitations underscore the value of AR-based systems, which embed guidance and feedback directly into the physical environment, enabling more natural interaction and contextual relevance.

%%%%% AR
\textbf{AR-based training systems}, in contrast, embed digital guidance into the real world, allowing users to stay aware of their surroundings while receiving visual or interactive cues. This context-aware augmentation supports safer, more natural practice, particularly for full-body and sport-specific movements. Several systems have leveraged AR to support physical skill learning.  
YouMove~\cite{anderson2013youmove} used a large-scale AR mirror to guide users through recorded movements, fading cues over time to encourage retention.  
Building on this idea, AR-Enhanced Workouts~\cite{wu2023ar} employed pose-based overlays for home fitness, showing that situated feedback improves understanding and real-time correction.  
Lin et al.~\cite{lin2021towards} visualized basketball shot trajectories in real time, enhancing shooting consistency and form awareness.  
Tai Chi AR Trainer~\cite{chen2020augmented} and Soccer MR Trainer~\cite{kim2022mixed} extended AR guidance to martial arts and outdoor sports, providing pose evaluation via head-mounted displays.  
In the context of team sports, VisCourt~\cite{cheng2024viscourt} delivered in-situ tactical instruction for basketball, enabling multiplayer coordination and decision-making in real physical environments. Despite their benefits, existing systems often lack personalization and rely on limited binary or part-level feedback, without leveraging AI to interpret nuanced motion data.  
In contrast, our system is designed to support more dynamic and personalized motor training by adapting to individual performance and delivering fine-grained, context-aware feedback through integrated spatio-temporal analysis and AI-powered interpretation.

\cglee{\subsection{Feedback Mechanisms for Embodied Motor Learning}}
\textbf{Visual Feedback for Human Motor Training.}
Visualization plays a key role in exploring human motion data, helping researchers understand movement patterns and communicate insights effectively.
When integrated into immersive environments, such visualizations enhance observation and deepen understanding of complex motion.
A common method is to visualize 3D trajectories of body parts (e.g., hands, head, feet) over time~\cite{kloiber2020immersive, hubenschmid2022relive}.
For instance, MIRIA~\cite{buschel2021miria} provides 3D trajectories, heatmaps, and scatterplots from recorded movements and interactions in mixed reality, supporting behavior analysis.
However, such visualizations often emphasize global position and overlook detailed posture, limiting contextual interpretation.
To address this, skeleton- or avatar-based visualizations have been adopted to more realistically represent posture and gesture~\cite{chan2007immersive, chan2010virtual, bernard2013motionexplorer, jang2015motionflow, kloiber2022immersive}, allowing users to observe motion with minimal cognitive load.
Wu et al.~\cite{wu2023ar} proposed a design space for visualizing workouts, categorizing visualizations by task, data type, and spatial relation to the body, and offering practical guidance for designing AR-based motion visualizations.

\cglee{\textbf{Embodied Skill Learning Beyond Sports.} From an embodied learning perspective, motor skill acquisition involves more than reducing kinematic error: learners gradually align internal sensations of balance, rhythm, timing, and weight transfer with external cues and instructional structures~\cite{bailey2010body, braun2022influence, wood2009tacit, leman2007embodied, juntunen2020ways}. 
HCI and the learning sciences have examined such embodied motor learning in domains beyond sports. 
VR and AR systems for dance and yoga use motion capture, avatar guidance, and overlaid videos to support timing, rhythm, and expressive movement flow, not only spatial accuracy~\cite{chan2010virtual, jo2023flowar}. 
In music education, embodied accounts of instrumental learning emphasize how students couple proprioceptive sensations with auditory outcomes and teacher feedback to refine technique~\cite{nijs2009musical, van2010musicjacket, leman2007embodied}. 
Work on traditional and hybrid craft similarly shows that expertise is transmitted through situated demonstrations, material engagement, and finely tuned bodily cues rather than purely verbal instructions~\cite{wood2009tacit, golsteijn2014hybrid}. 
% Building on these perspectives, we situate ViSTAR as a system that highlights embodied aspects of basketball skill learning by linking external feedback to how movements feel, while still operating on observable kinematic data. 
% Rather than directly measuring or training inner bodily sensations, the system provides multi-faceted feedback about posture, rhythm, and timing that athletes can use to reflect on their balance.
}
\cgleetwo{
Building on these perspectives, we situate ViSTAR as a system that highlights embodied aspects of basketball skill learning by linking external feedback to how movements feel, while still operating on observable kinematic data.
Concretely, ViSTAR translates kinematic deviations into feedback cues about posture, rhythm, and timing, anchored to drill phases and across-repetition comparisons, prompting athletes to self-check felt stability and weight transfer.
Rather than directly measuring or modeling internal sensations, ViSTAR uses these kinematics-derived cues to prompt athletes’ reflection on balance.
}

\subsection{AI Sports Coaching System}
AI is transforming the sports industry by enhancing performance analytics and decision-making~\cite{zhao2025survey, shih2017survey, wang2021research}.  
This work focuses on the role of AI in sports coaching, addressing the limitations of traditional coaching resources, which are often restricted by cost, location, or availability. To fill this gap, researchers have explored AI-driven systems that provide automated or semi-automated feedback during training.
Early systems were typically rule-based, relying on expert-encoded domain knowledge. For instance, Yin et al.~\cite{yin2000knowledge} proposed a knowledge-based framework for intelligent team training. While interpretable, such systems lacked adaptability and scalability.  
Recent advances have shifted toward data-driven approaches that use sensors and machine learning to analyze performance and deliver feedback. ARRow~\cite{iannucci2023arrow}, for example, is an AR system that provides real-time rowing feedback by visualizing biomechanical metrics such as stroke timing and posture using 3D skeletons.  
Similarly, PoseCoach~\cite{liu2022posecoach} offers customizable video-based coaching by comparing novice and expert poses, presenting differences through 3D animations. These systems generally use computer vision to compare user motion to ideal models and visualize discrepancies for correction. More recently, Large Language Models (LLMs) have enabled adaptive, conversational coaching. GPTCoach~\cite{jorke2024supporting}, for instance, combines wearable sensor data with motivational interviewing to deliver personalized activity plans.
Building on these developments, our system integrates LLM-powered feedback with immersive visualizations. 
Unlike prior systems that rely solely on either visual or textual cues, our system offers a tightly integrated approach that combines immersive 3D motion visualizations with AI-generated verbal feedback. 
The verbal feedback offers timely, specific, and actionable guidance, helping users recognize posture errors, understand corrections, and refine their movements more effectively during practice.

% This paper focuses specifically on AI applications in sports coaching—a domain where high-quality, personalized coaching remains limited. Traditional computer-assisted sports training systems were predominantly rule-based or heuristic. For instance, Yin et al.~\cite{...} developed a knowledge-based system for team training that relied on predefined models of ``correct'' technique.
% Modern approaches, however, have shifted toward data-driven methods. These systems leverage sensors and machine learning to provide adaptive feedback tailored to individual athletes. For example, systems like ARRow~\cite{...} and~PoseCoach \cite{...} utilize computer vision techniques to compare an athlete's motion with an idealized benchmark, subsequently visualizing the differences to facilitate improvement.
% More recently, advances in large language models (LLMs) have introduced the potential for interactive and personalized coaching. GPTCoach \cite{...} exemplifies this trend, offering natural language-based feedback that adapts to the athlete's context. Building on these developments, our work aims to deliver intelligent, human-understandable feedback through immersive interfaces that align with the spatial dynamics of sports training.

%% file: Source/3_formative_study.tex
\section{FORMATIVE STUDY}
We conducted a formative study to understand the current practices and challenges players encounter 
when learning and practicing basketball skills.

\subsection{Study Setup}

\textbf{Participants:} 
We interviewed five university basketball players (mean age: 20.8 years, mean playing experience: 13.2 years) and two basketball coaches (mean age: 25.5 years, mean coaching experience: 5.5 years) to gain insights into their experiences and perspectives. 

\textbf{Procedure:} 
Each session consisted of a 45-minute initial interview, a 15-minute skill training demonstration, and a 10-minute final interview. With consent to record, we asked participants about their background, training routines, self-learning methods, and challenges. 
After the interview, we showed three prepared skill training videos and asked participants to choose one to learn. They then practiced the skill while reviewing their captured motions. 
Finally, we conducted a post-training interview to understand their experience and how their perspectives evolved.
% and to gather insights into their experiences.  

\textbf{Data Analysis:}
We transcribed the audio recordings and applied Grounded Theory analysis~\cite{muller2012grounded, muller2014curiosity} to identify key difficulties. 
Two authors independently coded the transcripts and jointly reviewed the codes to establish categories. 
In cases of disagreement, the coders resolved differences through discussion. Inter-coder agreement, measured using Cohen's Kappa, was 0.90.

\subsection{Findings Summary}
The interviews revealed that participants' skill-training process typically consists of three major components:  
1) \emph{Learning}: repeatedly watching video clips;
2) \emph{Practices}: attempting to imitate the demonstrated movements;
\cglee{
3) \emph{Improving}: recording and reviewing their own motions using tools such as loop playback and slow motion, then making adjustments based on perceived errors and how the motion felt. 
However, participants often struggled to translate a vague sense that something was “off”—for example, in their balance, timing, rhythm, or weight transfer—into concrete, body-part–specific corrections. 
We also identified three major difficulties in participants' skill-learning process:
}

% 3) \emph{Improving}: 
% recording and reviewing their own motions,  using tools such as loop playback and slow motion, and  making adjustments based on perceived errors.  
% We also identified three major difficulties in participants' skill-learning process:

\textbf{D1. Difficulty Understanding Skills from External Resources.}  
Video clips were the primary resource for learning, but participants noted that such materials are limited in perspective and lack multi-angle views.  
As one participant observed, \textit{``Being able to pause and look at the play from multiple angles would be extremely helpful.''}  
Without varied viewpoints or 3D representations, participants struggled to perceive spatial structures, body alignment, timing, and coordination~\cite{martinez2022comparison}.  
Participants also reported that passive observation alone was insufficient, particularly for complex motor skills. As another remarked, \textit{``It’s easier to understand when someone demonstrates in front of me.''}  
These findings highlight the need for guided instruction that breaks motions into smaller, digestible segments and emphasizes subtle but critical details such as joint articulation and timing accuracy.

\textbf{D2. Difficulty Recognizing Errors During Practice.}  
Many participants expressed a desire to compare their performance directly, side-by-side, with an expert model.  
One participant explained, \textit{``Comparing my shot side-by-side with another could be insightful, as it would help me see the coordination of movements more clearly.''}  
Current tools, however, do not support synchronized comparisons.  
As a result, participants find it easier to detect mistakes when receiving external feedback. As one admitted, \textit{``It’s hard to judge on my own. Having a more objective way to spot mistakes helps me understand better''}~\cite{dunning2004flawed}.  
% While some participants clip key moments to review with coaches, this support is not always available, leaving athletes to struggle with inefficient self-assessment.  
\cglee{Without such references, their internal sense of whether a movement was “right” or “wrong” often remained vague.}  
While some participants clip key moments to review with coaches, this support is not always available, leaving athletes to struggle with inefficient self-assessment.

\cglee{\textbf{D3. Difficulty Turning Intuition into Actionable Corrections.}}
Even when participants sensed that a movement felt wrong, they lacked the guidance to correct it.  
\cglee{Participants often described how the move felt overall (e.g., off-balance, out of rhythm, or mistimed), but still struggled to pinpoint which body part to adjust, in what direction, and at what moment, particularly without expert input.}
% They struggled to determine which body part to adjust, in what direction, and at what moment, particularly without expert input. 
% This underscores the importance of specific, actionable feedback for effective self-correction.  
As one participant explained, \textit{``I couldn’t determine which parts I needed to improve, leaving me feeling uncertain.''}  
% This underscores the importance of specific, actionable feedback for effective self-correction.  
Prior work has shown that high-quality feedback boosts motivation, self-confidence, and training satisfaction~\cite{carpentier2016predicting}.  
Yet feedback quality varies with a coach's expertise and communication style, and coaches are not always present.  
Participants noted that vague comments such as \textit{``Be more aggressive''} left them confused, 
whereas precise instructions like \textit{``Shift your weight to the right foot''} were more helpful.  
% These findings underscore the need for systems that can deliver clear, consistent, and context-specific feedback, independent of time, location, or coach availability.
\cglee{Overall, these findings point to the need for systems that can deliver clear, consistent, and context-specific feedback that connects how a move feels (e.g., balance, rhythm, timing, weight transfer) with concrete, interpretable cues about how to adjust one’s posture or coordination, independent of time, location, or coach availability.}

%% file: Source/4_design_consideration.tex
\section{ViSTAR: AR Coaching Agent for Behavioral Skills Training}
\label{sec:design}

\begin{figure*}[t]
    \centering
    \includegraphics[width=1.0\linewidth]{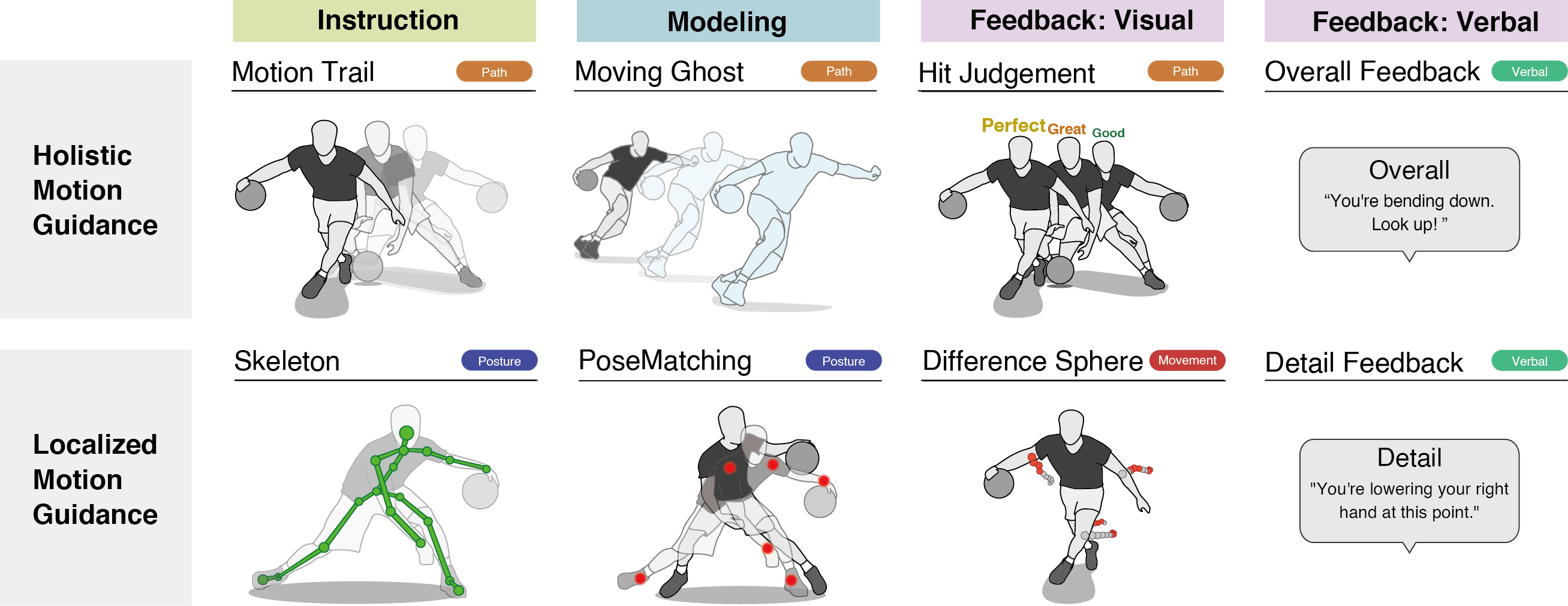}
    \caption{Motion guidance strategies across BST stages. The top row represents Holistic Motion Guidance (e.g., Motion Trail, Moving Ghost, and Hit Judgement) that supports overall flow and coordination through path guidance. The bottom row shows Localized Motion Guidance (e.g., Skeleton, PoseMatching, Difference Sphere) focusing on joint-level feedback for fine-grained correction.}
    \Description{Figure 2: Comparison of motion guidance strategies across training stages. The figure is organized into two rows, with the top row showing holistic guidance techniques that emphasize overall movement flow using trajectory-based visual cues, and the bottom row showing localized guidance techniques that provide joint-level feedback through skeletal and difference-based visualizations.}
    \label{fig:bst}
    \vspace{-2em}
\end{figure*}

% \cglee{Movement Guidance \& overall and detail - formative study (what I should do, x know improve), - design (gap found - overall and detail), - result (highlight - benefits detail driven visualization)}

To address the difficulties identified in our formative study, 
we designed \namespace, a AR system based on BST framework to provide both visual and verbal guidances to help users in basketball skill learning.

%%% General usage pipeline, talk about how this pipeline address D1, D2, and D3.
\subsection{Usage Scenario}
To illustrate the workflow of \namespace, we describe a hypothetical learner, Kooto, practicing a crossover dribble. First, \namespace presents the target motion as a 3D AR avatar, segmented into phases and viewable from multiple angles. Unlike conventional videos that restrict learners to a fixed viewpoint, the 3D AR avatar in \namespace can be viewed from multiple angles. Kooto can pause, rotate, and focus on subtle details such as joint articulation and timing. This flexibility enables him to examine the motion from different perspectives and gain a clearer understanding of complex movements, thereby addressing D1 \cglee{by offering richer visual access to the motion.  }
As he follows along, Kooto’s 3D pose is captured in real time and automatically aligned with the reference model. These visual overlays and synchronized side-by-side comparisons offer a consistent reference for self-assessment, helping Kooto observe subtle differences that can be overlooked, \cglee{helping to mitigate D2 by making discrepancies more visually salient.} %thereby addressing D2.
Kooto then rehearses the skill by recording his own attempts, replaying them with expert overlays, and iteratively refining his motion.  
Finally, \namespace provides comprehensive multi-faceted feedback: visual heat maps emphasize regions where adjustments are beneficial, and verbal coaching offers precise and actionable suggestions such as ``shift your weight to the right foot before crossing over.'' \cglee{This specificity aims to move beyond vague remarks (e.g., “be more aggressive”) by offering more targeted guidance that players can adapt to their own body and style, thus aiming to address D3 by offering more specific, coach-like cues.}
% This specificity avoids vague remarks and instead provides targeted guidance that learners can adapt to their own body and style, \cglee{thus aiming to address D3 by offering more specific, coach-like cues.} %thus addressing D3. 
Through this workflow, \namespace operationalizes the Behavioral Skills Training (BST) framework of instruction, modeling, rehearsal, and feedback in an immersive AR environment.

% To illustrate the workflow of \namespace, we describe a hypothetical user, Kooto, learning a new skill.  
% \zt{please refine the text below to tell a complete story. can use the one in the video demo. maybe can refer to teaser}
% First, \namespace displays the target motion in 3D AR and breaks it down into key steps, directly addressing D1 by offering multi-angle, segmented demonstrations.  
% The user then follows along, while \namespace automatically captures their 3D pose using an external camera, compares it with a reference model, and highlights discrepancies (D2). Finally, after recording a completed attempt, the system generates explainable, comprehensive feedback in both visual and verbal forms (D3).  

% \begin{figure*}[t]
%     \centering
%     \includegraphics[width=1.0\linewidth]{images/Figure_Implement.eps}
%     \caption{Motion guidance strategies across BST stages. The top row represents Holistic Motion Guidance (e.g., Motion Trail, Moving Ghost, and Hit Judgement) that supports overall flow and coordination through path guidance. The bottom row shows Localized Motion Guidance (e.g., Skeleton, PoseMatching, Difference Sphere) focusing on joint-level feedback for fine-grained correction.}
%     \label{fig:bst}
% \end{figure*}

\begin{figure*}[t]
    \centering
    \includegraphics[width=1.0\linewidth]{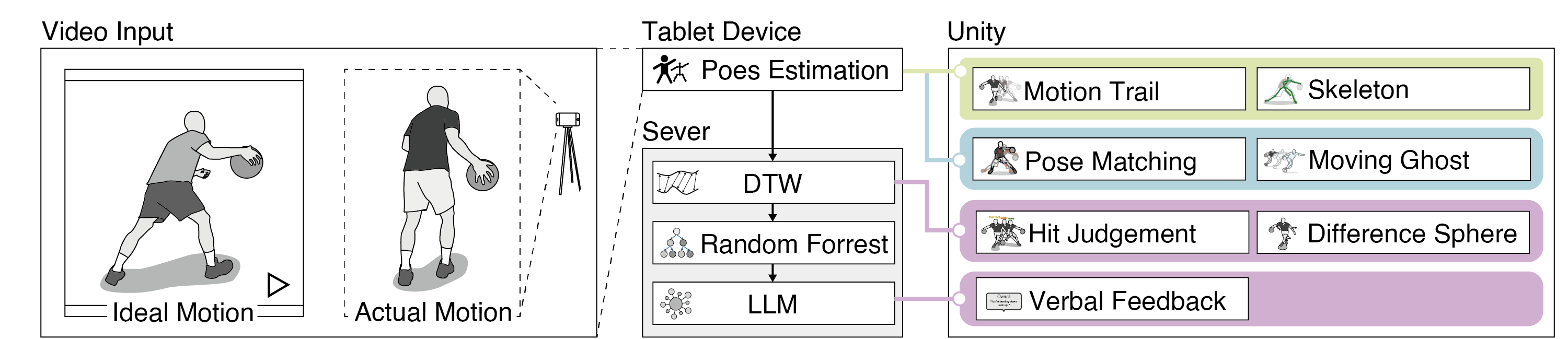}
    \caption{Overview of the system workflow. User motion is analyzed using pose estimation, DTW, and Random Forest, and the resulting analysis is used to generate motion guidances, which are visualized in Unity through multi-faceted feedback.}
    \Description{Figure 3: Overview of the system workflow. User motion is processed through pose estimation and temporal alignment, followed by analysis to generate motion guidance. The resulting feedback is visualized in an augmented reality environment using a game engine.}
    \label{fig:system}
\end{figure*}

\subsection{BST-grounded Guidance Design}
The design of \namespace is grounded in the Behavioral Skills Training (BST) framework (\autoref{fig:bst}), a well-established method for improving skill performance across sports domains~\cite{sarokoff2004effects, wiley2024behavioral, harris2020using, o2020effect}.  
BST consists of four stages:  
1) \emph{Instruction} — the trainer clearly describes the target skill,  
2) \emph{Modeling} — the trainer demonstrates the skill in detail,  
3) \emph{Rehearsal} — the learner practices the skill, and  
4) \emph{Feedback} — the trainer reviews the learner’s performance and suggests improvements.   
\cglee{Beyond structuring error recognition, this framework also highlights embodied aspects of skill learning by organizing how learners move between attending to the overall feel of a movement (e.g., rhythm and timing) and making specific postural adjustments, which informed our guidance design.}

In our design, \namespace acts as \textbf{the trainer} by providing guidance in the \emph{Instruction}, \emph{Modeling}, and \emph{Feedback} stages.  
To achieve this, we incorporated motion-guidance strategies identified in prior work~\cite{elsayed2022understanding}, spanning two complementary dimensions:  
\textbf{Holistic guidance}, which emphasizes overall flow, timing, and full-body coordination \cglee{(and can help learners notice changes in their balance across the move)}, directing users along intended trajectories; 
\textbf{Localized guidance}, which emphasizes accuracy in specific joints or body segments, such as highlighting key-frame poses to reinforce posture while accounting for individual differences (e.g., height, limb-to-torso ratio). 
We instantiated these strategies into guidance features for each stage as follows:

\begin{itemize}
    \item \textbf{Instruction Guidance:}  
    Skills are introduced through motion trails that visualize trajectories, combined with a 3D avatar and overlaid skeleton that highlights joint articulation (\autoref{fig:bst}a). This provides a clear reference of spatial and temporal structures, addressing D1.

    \item \textbf{Modeling Guidance:}  
    Expert motions are demonstrated using moving ghost and a pose-matching interface (\autoref{fig:bst}b).  
    The interface segments skills into discrete poses, providing further support for D1.
    Moreover, when the user follow the discrete poses, the system can provide real-time feedback.
    Specifically, when misalignment occurs, red spheres appear on the relevant joints and disappear once corrected, offering progressive, joint-level feedback on angles and timing. This partially addresses the error recognition challenges (D2).

    \item \textbf{Feedback Guidance (Visual):}  
    In the feedback stage, \namespace combines holistic and localized guidance (\autoref{fig:bst}c) to address D3.  
    For holistic guidance, a hit-judgment mechanism provides rhythm-game-style cues — ``Perfect,'' ``Excellent,'' ``Great,'' ``Good,'' or ``Imprecise'' — to reflect temporal accuracy. This helps users refine timing and coordination according to their physical characteristics and preferred tempo.  
    For localized guidance, \namespace synchronizes the learner’s motion with an expert reference in a side-by-side AR view. Misaligned joints are marked with red spheres that fade as alignment improves, providing immediate, interpretable feedback on spatial deviations.  
    Together, these two visual feedback modes help learners understand both \emph{when} and \emph{where} their execution diverges from the reference.  

    \item \textbf{Feedback Guidance (Verbal):}  
    To further address the lack of actionable coaching, \namespace converts joint-level analysis into natural language suggestions (\autoref{fig:bst}d).  
    These context-specific prompts mirror the effectiveness of precise coaching comments observed in our formative study (e.g., ``Shift your weight to the right foot''), enabling iterative refinement during solo practice without requiring a coach to be present.  

\end{itemize}

%% file: Source/5_system.tex
\section{Technical Implementations}
\label{sec:system}

\namespace leverages AR's spatial visualization and embodied interaction capabilities, together with the language generation capabilities of LLMs, to support the guidance introduced in Sec.~\ref{sec:design}. 
Here, we detail the key technical details and innovations for each stage. 
\cglee{\namespace focuses on observable kinematics (e.g., posture, timing, and rhythm) and turns them into spatial and verbal cues that learners can align with their own bodily sensations. Our goal is not to fully model embodied experience, but to surface when balance, weight transfer, or tempo likely feel “off” and translate these patterns into concrete, coach-like guidance. To keep the hardware resource low and make deployment feasible, \namespace relies on a single front-facing RGB camera for 3D reconstruction. This design avoids the cost and calibration overhead of multi-camera setups. However, a multi-camera configuration could further improve joint reconstruction quality and robustness. }

\subsection{Instruction -- 3D Reconstruction and Animate Avatar}
% \zt{Talk about the 3D reconstruction. how we reconstruct a 2D video into 3D instruction and display in AR to the user.}

To support the instruction phase and address D1, we integrate a monocular pose estimation model~\cite{aoyagi2022development} that reconstructs full-body motion from a single 2D video stream. The model predicts 3D joint locations and rotation parameters from RGB frames, which we then retarget to a rigged avatar. This process transforms ordinary camera input (e.g., from a smartphone) into a 3D representation of expert demonstrations as well as the user’s own performance. Users can select virtual coaches (e.g., their favorite players), which serve as consistent demonstrators. Using joint rotation data extracted with the model~\cite{aoyagi2022development}, we animate a 3D avatar that is viewable from any angle, reducing spatial ambiguity. To support understanding, the system overlays \textit{motion trails} (path guidance) and \textit{skeleton visualization} (posture guidance), as illustrated in \autoref{fig:bst}. Playback controls (pause, rewind, slow) let users inspect tempo and detail, revealing subtle joint and timing cues.
% \cglee{Beyond visual clarity, being able to walk around the avatar and inspect motion trails from different viewpoints helped players relate what they saw to how their own balance and weight transfer felt in specific phases of the move, as we report in Sec.~\ref{sec:results}.}

\subsection{Modeling -- Motion Step-by-Step Breakdown}
\label{sec:posetrack}
% \zt{Talk about how you support the modeling stage. btw, have you used DTW in the real-time feedback?}

To support skill understanding (D1), we implement a dynamic ``moving ghost'' visualization in the AR scene. The ghost is realized by animating a semi-transparent  avatar with expert joint rotations for each frame. This path guidance goes beyond static frames, presenting the entire movement fluidly in time and space as a holistic guidance.  For pose matching, \namespace normalizes joint positions by user-reported height so that expert and learner poses are represented in a common reference scale. 
% Pose matching operates over segmented intervals (4 or 8 segments per skill) using both joint angles and positions. For each joint, we compute the angular deviation between the expert’s and learner’s rotation matrices, and the positional error as the Euclidean distance between normalized joint positions. A joint is considered aligned when the angular error is below $30^\circ$ and the positional error is below $0.1\,\mathrm{m}$. When at least 75\% of joints in a segment meet these criteria, the system advances to the next segment. 
\cglee{Pose matching operates over segmented intervals (4 or 8 segments per skill) using both joint angles and positions.} For each joint, we compute the angular deviation between the expert’s and learner’s rotation matrices, and the positional error as the Euclidean distance between normalized joint positions. A joint is considered aligned when the angular error is below $30^\circ$ and the positional error is below $0.1\,\mathrm{m}$, providing a coarse tolerance band that leaves room for individual rhythm while still flagging clearly problematic deviations. When at least 75\% of joints in a segment meet these criteria, the system advances to the next segment.\cglee{ We do not perform a temporal alignment over the entire sequence, since the expert avatar already leads the learner through a scripted sequence of 4 or 8 segments.}
Real-time feedback is provided by overlaying red spheres on joints that exceed these thresholds.  Markers fade as alignment improves, providing continuous correction cues. Crucially, these criteria do not enforce perfect replication, but promote adaptation: players internalize essential movement patterns and refine them to suit their own body characteristics and physical capabilities (e.g., anthropometrics, mobility, strength), focusing on effective, sustainable execution rather than identical form.

% To help users understand skills (D1), we implement a dynamic ``moving ghost'' visualization: a semi-transparent animated overlay of the expert avatar’s motion. This path guidance goes beyond static frames, presenting the entire movement fluidly in time and space as a holistic guidance. 
% To account for differences in body proportions, \namespace asks users to input their height, using to normalize joint positions during pose comparison.
% Pose matching uses joint positions and rotation angles for each temporal segment. Each skill is split into 4 or 8 segments for localized guidance. The system overlays red spheres in real-time on joints that deviate from the expert pose in \autoref{fig:bst}.  
% A joint is considered matched when normalized angular and positional errors fall below $30^\circ$ and $0.1,\mathrm{m}$. When at least $75\%$ of joints in a segment match, the system advances to the next pose. 
% Markers fade as alignment improves, providing continuous correction cues. Crucially, these criteria do not enforce perfect replication, but promote adaptation: players internalize essential movement patterns and refine them to suit their own body characteristics and physical capabilities (e.g., anthropometrics, mobility, strength), focusing on effective, sustainable execution rather than identical form.

% angle 30 degree
% distance 0.1m
% Match joints more than 75% of all joints

\subsection{Feedback -- Generating Multi-faceted Feedback}

\cglee{Unlike the segmented pose matching (\autoref{sec:posetrack}), where the learner rehearses one phase at a time under system-guided timing, the multi-faceted feedback operates on the full skill performed at the learner’s own pace. These recordings typically contain local tempo variations within the skill as well as extra frames before and after the core movement, so we use a DTW-based temporal alignment to fairly compare expert and learner trajectories and trim onset/offset noise over the entire sequence.}
We describe how \namespace implements multi-faceted feedback through hit judgement, spatial visual feedback, and verbal feedback.  
Once the video is captured, \namespace reconstructs the player’s 3D movement and uploads it to the server, where Dynamic Time Warping (DTW)~\cite{salvador2007toward} and motion alignment algorithms compute temporal and spatial correspondences with the expert reference (\autoref{fig:system}). This alignment ensures that feedback remains consistent even when the learner’s execution differs in tempo or scale.

\subsubsection{Motion Alignment via Sliding Window Matching}
Since users manually start/stop recording, sequences include extraneous frames (e.g., button taps). We remove these irrelevant segments with a sliding window over the full sequence: for each window, we compute the DTW distance to the full expert motion and select the window with the minimum distance as the performed segment. This jointly optimizes start and length, making the method robust to speed variation:

\[
t^* = \argmin\limits_{t \in [0, L_u - L_i]} \text{DTW}(M_{\text{user}}[t:t+L_r], M_{\text{ref}})
\]

\noindent
where \( L_i \) is the length of the ideal motion and \( L_u \) is the length of the user's motion.

Furthermore, due to variations in player speed, the duration of user motions can be longer or shorter than the ideal. Using the best-aligned start index \( t^* \), we perform variable-length matching by adjusting the window length \( L_w \) in the range \( [0.5 L_i, L_u - t^*] \) and compute:
\[
L_w^* = \argmin\limits_{L_w} \text{DTW}(M_{\text{user}}[t^*:t^* + L_w], M_{\text{ref}})
\]
This approach ensures robustness to speed variations by selecting the best-aligned user segment based on both starting point and optimal length. \cglee{All subsequent multi-faceted feedback (for hit judgement, difference sphere, and verbal feedback) are applied on the DTW-aligned trajectories.}

\subsubsection{Hit Judgement and Spatial Visual feedback}
\label{sec:visualfeedback}
Temporal accuracy, timing and rhythm, is another critical yet hard to self-assess (D2). As a holistic guidance, our system implements a hit judgment mechanism over segmented intervals. The expert motion is divided into equal parts (e.g., 4 or 8 segments), and each segment is compared to the corresponding user motion by measuring its duration. \cglee{Similar uniform time-normalization is widely used in biomechanics~\cite{dicesare2020high, zhu2023development}, where movement cycles are rescaled to a common duration before comparing joint kinematics or other performance measures across trials.} Timing accuracy is computed using the following formula:
\[
\text{Score} = \max\left(0,\, 100 - \left| \frac{T_\text{ideal} - T_\text{actual}}{T_\text{ideal}} \right| \times 100 \right)
\]
Scores map to visual labels—\textit{Perfect} ($\ge90\%$), \textit{Excellent} ($\ge80\%$), \textit{Great} ($\ge70\%$), \textit{Good} ($\ge60\%$), \textit{Imprecise} ($\ge50\%$), providing immediate cues on rhythm and speed for coach-free self-correction.

For spatial feedback, we render user and expert side-by-side in AR with synchronized playback. 
\cglee{Red spheres highlight joints whose DTW-aligned positional error exceeds a distance threshold (we use the same $0.1,\mathrm{m}$ tolerance as in the segmented phase), while gray ones denote joints within tolerance.}
% Red spheres highlight joints exceeding a distance threshold, while gray ones present aligned joints. Spheres update frame by frame, showing 
\emph{when} and \emph{where} deviations occur. Combined with segment-level timing scores, this dual-layer feedback offers interpretable spatiotemporal cues that enable self-directed review and correction.

\subsubsection{Generating Verbal Feedback}
\label{sec:verbalfeedback}
To help users identify and correct mistakes (D3), \namespace provides actionable verbal feedback. We convert both ideal and user numerical motion data into textual inputs for LLMs, which are not designed to handle raw joint angles. Without contextualization, LLMs struggle to interpret such data \cite{levy2024language}. While some approaches leverage time-series structures \cite{jin2023time}, they require large-scale training. Instead, \namespace proposes a lightweight method combining dynamic time warping and random forest to identify and summarize key joint differences into concise textual feedback.

\begin{figure}[t]
    \centering
    \includegraphics[width=1.0\linewidth]{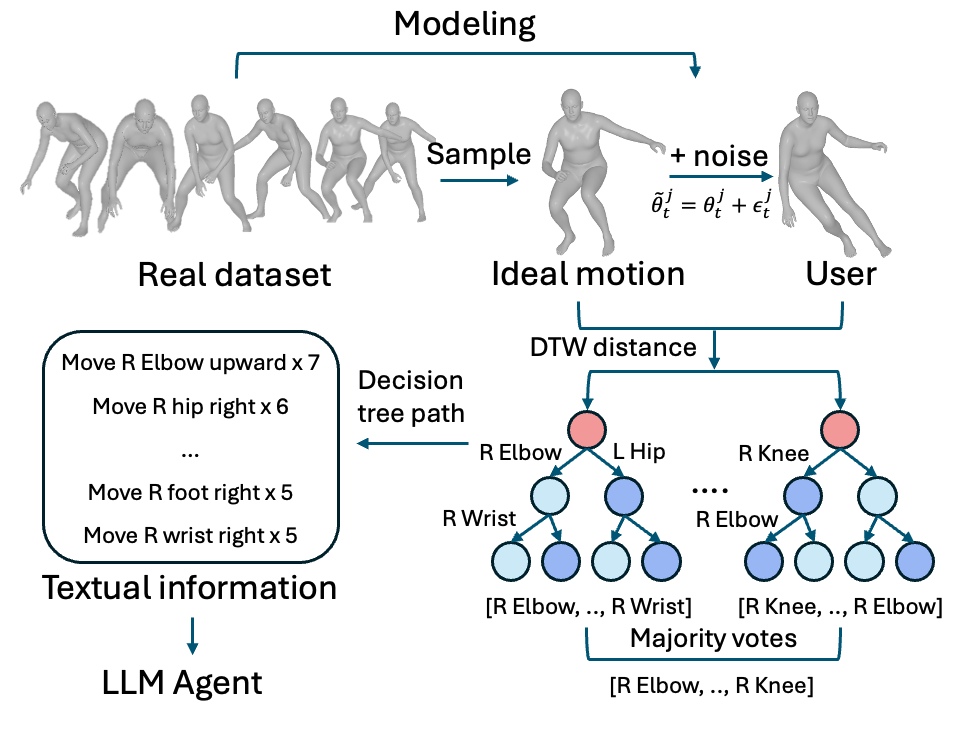}
    \caption{Verbal feedback generation pipeline. User motion is compared to ideal motion using DTW and a decision tree, and key misaligned joints are identified. An LLM generates feedback based on motion descriptors from the dataset.}
    \Description{Figure 4: Verbal feedback generation pipeline. The figure shows how user motion is compared with an ideal reference to identify misaligned joints, which are then used to generate verbal feedback through a language-based model.}
    \vspace{-0.5cm}
    \label{fig:example}
\end{figure}

\begin{figure*}[t]
    \vspace{-0.4cm}
    \centering
    \includegraphics[width=0.9\linewidth]{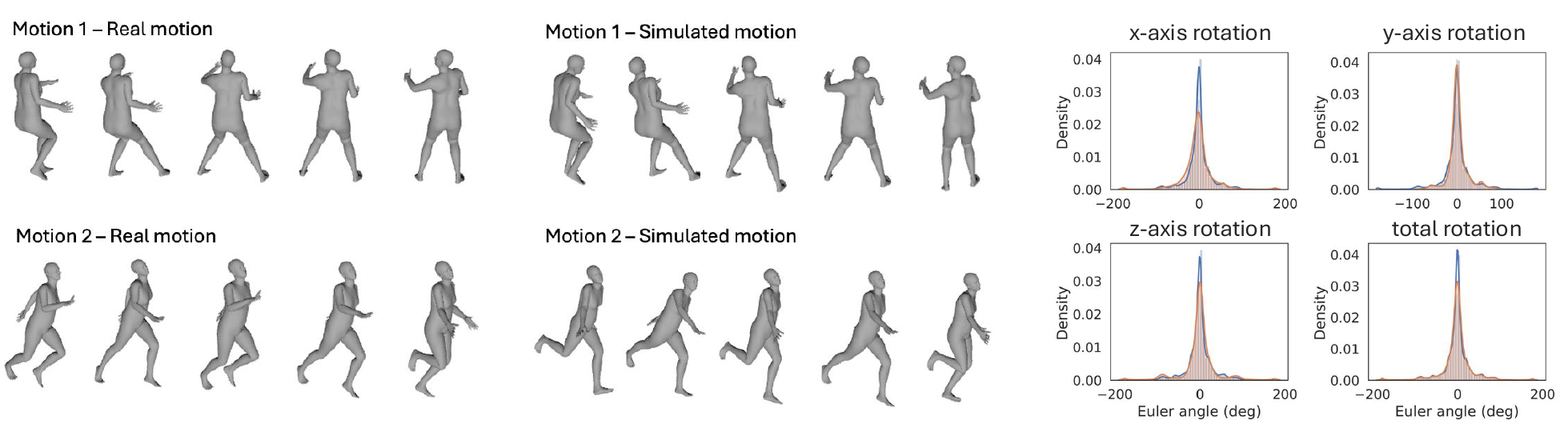}
    \caption{Comparison between real and simulated motions, along with distribution visualizations.}
    \Description{Figure 5: Visual comparison between real user motion and simulated motion produced by the virtual user motion generation pipeline.}
    \label{fig:simulation}
    \vspace{-0.4cm}
\end{figure*}

\textbf{Identifying different joints.}
Euclidean distance is ill-suited for spatiotemporal motion. We compare Skinned Multi-Person Linear model (SMPL)~\cite{loper2023smpl}-based joint-rotation sequences ($\text{frames}\times24\times3$) using FastDTW \cite{salvador2007toward}. When a joint–angle pair exceeds a threshold, it is verbalized (e.g., ``move your left knee rightward; reduce excessive motion''), aggregated, and passed to an LLM to produce natural-language feedback. This naive listing can still surface too many joints, risking overload and diluting focus.

\textbf{Summarizing Important Motions using Random Forest.}
To reduce excessive feedback, we use a Random Forest classifier \cite{breiman2001random} to prioritize the most critical joint movements, focusing on informative and actionable cues rather than all joints.  Random Forest, an ensemble method of decision trees, offers interpretable decision paths. By analyzing these paths (not model-level feature importance), we can identify which joint-angle differences most influenced the prediction, enabling instance-level interpretability. We selected Random Forest over deep neural networks or XGBoost \cite{chen2016xgboost}  due to its transparency, lightweight nature, and robustness with limited data. Although XGBoost supports instance (e.g., level interpretation via SHAP values), SHAP \cite{lundberg2017unified} is computationally expensive and less intuitive to trace. Random Forest offers easier integration with LLMs and requires minimal tuning.
\cglee{We de-emphasize spine and torso joints in verbal feedback, since trunk posture and balance are easily visible in the full-body AR avatar and joint markers, whereas multi-limb coordination errors are harder to spot without explicit guidance. In line with prior systems that focus metrics on a set of limb joints~\cite{elsayed2022understanding,debarba2018augmented}, our LLM prompts therefore primarily summarize recommendations at the multi-limb (arm/leg) level, while still surfacing spine-related cues when they are primary sources of error.}

\textbf{Training with a Synthesized Dataset.}
Since no dataset pairs expert and user motions for the same action, we synthesize one from public mocap (e.g., CMU MoCap~\cite{cmu_mocap}). We extracted 61 basketball-related motions (e.g., dribbling, shooting) and segmented them into 178 shorter sequences for finer variationthen split $80\%$/$20\%$ for train/test. To simulate realistic errors, we added joint-level noise by computing frame-wise joint angle differences across time \( t \), joint \( j \), and axis \( a \in \{x, y, z\} \), then sampling from noise distributions to create plausible user motion variants (see \autoref{fig:simulation}).

% total num / acc 

% This synthetic data generation pipeline is generalizable and applicable to other motion datasets or domains. 
In principle, the same synthetic perturbation strategy could be adapted to other motion datasets, but in this work we only apply and validate it on basketball-related skills.
While generative models such as Motion Diffuse \cite{zhang2024motiondiffuse} can create motion variations, they have limited capacity to model the inter-personal variations for the same action Instead, we model temporal differences in joint angles by fitting an exponential distribution to their absolute values. 
For each joint $j$ and rotation axis $a$, we compute the temporal angle differences as:
\[
\Delta \theta_t^{j,a} = \theta_t^{j,a} - \theta_{t-1}^{j,a}, \quad t = 2, \ldots, T,
\]
and fit an exponential distribution $\mathcal{E}(\lambda_{j,a})$ to the set
\[
\left\{ |\Delta \theta_t^{j,a}| \right\}_{t=2}^T \sim \mathcal{E}(\lambda_{j,a}).
\]
This formulation captures how frequently and abruptly each joint tends to move over time. The parameter $\lambda_{j,a}$ is estimated from the entire sequence, reflecting the typical magnitude of motion for the given joint and axis. Using the fitted distribution, we generate synthetic noise for each joint $j$, axis $a$, and time $t$:
$
\epsilon_t^{j,a} =  \epsilon_{\text{exp}}^{j,a} + \epsilon_{\text{norm}}^{j,a}
\quad \text{where} \quad
\begin{cases}
\epsilon_{\text{exp}}^{j,a} \sim \text{Exponential}(\lambda_{j,a}) \\
\epsilon_{\text{norm}}^{j,a} \sim \mathcal{N}(0, \sigma_{j,a}^2)
\end{cases}
$
The exponential component $\epsilon_{\text{exp}}^{j,a}$ captures the typical magnitude and abruptness of joint movement, as estimated from the data. To better simulate the natural variability and randomness observed in human motion, we add a small amount of Gaussian noise $\epsilon_{\text{norm}}^{j,a}$, which introduces fine-grained stochasticity and avoids overly deterministic trajectories. Then, we create the noisy (user-like) motion:
$
\tilde{\theta}_t^{j,a} = \theta_t^{j,a} + \epsilon_t^{j,a}
$
We define the input feature vector \( \mathbf{x}_t \in \mathbb{R}^{N \times 3} \) as the difference between the noisy and original motion:
$
\mathbf{x}_t = \tilde{\theta}_t - \theta_t
$

where \( N \) is the number of joints (e.g., 24), and each joint has 3 rotational angles. 
% The corresponding label vector \( \mathbf{y}_t \in \{0, 1\}^N \) indicates which joints have been perturbed:
% \[
% \mathbf{y}_t^j =
% \begin{cases}
% 1, & \text{if noise is added to joint } j \\
% 0, & \text{otherwise}
% \end{cases}
% \]
The corresponding label vector $ \mathbf{y}_t \in {0, 1}^N $ indicates which joints have been perturbed, where the $j$-th component $y_t^j \in {0, 1}$ is $1$ if noise is added to joint $j$ at time $t$ and $0$ otherwise.
The synthetic dataset in \autoref{fig:example} provides training pairs $(\mathbf{x}_t,\mathbf{y}_t)$ for the Random Forest, where $\mathbf{x}_t$ contains motion features from both the original and noise-perturbed sequences as inputs, and $\mathbf{y}_t$ labels the joints to which noise was applied. 
% We ran an ablation over \texttt{n\_estimators} and \texttt{max\_depth} (Table~\ref{tab:ablation}). The best setting (\texttt{n\_estimators}=5, \texttt{max\_depth}=\texttt{None}) achieved average precision $0.8328$, recall $0.9496$, and F1 $0.8874$, and we adopt this configuration.
We extract instance-level explanations by tracing the decision paths of the trained forest and summarize them as inputs to the LLM for verbal feedback as shown in \autoref{prompts}. In decision tree path, each node encodes explicit thresholds (e.g., “joint angle > value”), aligning with our goal of generating interpretable, condition-driven feedback. In contrast, SHAP requires extra processing to convert scores into language, making it less suitable for fast, lightweight feedback generation.

\subsection{Implementation}
\label{sec:implementation}

We implemented four sequential steps: \textit{Instruction} (3D avatar), \textit{Modeling} (pose matching), \textit{Rehearsal} (practice/recording), and \textit{Feedback} (visual, timing, verbal). The system is built in \textit{Unity} and deployed on a \textit{Magic Leap AR headset} for immersive in-situ visualization. On the client side, Unity handles real-time rendering of 3D avatars, motion overlays, and AR feedback. 
For backend services, we use \textit{FastAPI} to manage data processing pipelines and communication between the AR client and external servers. Motion data is processed by an efficient pose estimation model~\cite{aoyagi2022development} that runs at 30 FPS on \textit{iPhone 16} for real-time capture, while higher-accuracy models can be integrated if latency is not critical. The verbal feedback module leverages the \textit{OpenAI GPT-4 API}, with pre-processed motion descriptors and Random Forest decision paths provided as inputs to generate concise coaching feedback.

\begin{promptbox}{Prompt for Generating Feedback Summarization}
\phantomsection\label{prompts}

You are an expert movement coach that helps users correct their body posture based on feedback from a motion analysis system. Your task is to take a list of movement instructions and summarize them in a \textbf{clear, natural, and structured} way that is easy for users to understand, while prioritizing the most important corrections.

\textbf{Instructions:}
\begin{enumerate}
    \item \textbf{Group movement instructions by body region:}
    \begin{itemize}
        \item \textbf{Upper Body}: shoulder, elbow, wrist $\rightarrow$ refer to as \textbf{arm}
        \item \textbf{Lower Body}: hip, knee, ankle, foot $\rightarrow$ refer to as \textbf{leg}
        \item Ignore spine/torso feedback unless it's the only area mentioned
    \end{itemize}

    \item \textbf{Identify the most critical issue} in each region:
    \begin{itemize}
        \item Prioritize joints with \textbf{multiple or strong directional issues}
        \item Focus on \textbf{excessive movement} that impacts balance or coordination
    \end{itemize}

    \item \textbf{Summarize feedback using only 1--2 concise sentences total:}
    \begin{itemize}
        \item Mention only \textbf{one or two joints} maximum
        \item Use language that is \textbf{specific}, \textbf{actionable}, and \textbf{understandable}
    \end{itemize}

    \item \textbf{Use user-friendly phrasing and avoid technical language:}
    \begin{itemize}
        \item Instead of ``Reduce movement right'', say ``Try to limit movement to the right''
        \item Instead of ``Maintain movement'', say ``Keep the movement as it is''
    \end{itemize}
\end{enumerate}

\textbf{Output Format:} Write a short paragraph that:
\begin{itemize}
    \item Clearly describes the most important movement issues
    \item Includes a \textbf{suggestion for improvement}
    \item Avoids overloading the user with too many joint names or directions
\end{itemize}

\textbf{Example Input:}
\begin{verbatim}
Move your left knee left and Reduce movement.  
Move your left elbow down and Reduce movement. 
\end{verbatim}

\textbf{Example Output:}
\begin{itemize}
    \item You may lose balance when lowering your left leg. Try adjusting your right leg as well.
    \item Your leg is moving in opposite directions. Try referencing a natural dribbling motion.
\end{itemize}

\end{promptbox}

\begin{table*}[t]
\centering
\caption{Ablation Study on \texttt{n\_estimators} and \texttt{max\_depth} for Random Forest. ``\textemdash'' indicates that \texttt{max\_depth} is set to None, meaning that the tree depth is unlimited.}
% \resizebox{\linewidth}{!}{%
\begin{tabular}{cccccc}
\toprule
\textbf{n\_estimators} & \textbf{max\_depth} & \textbf{Accuracy} & \textbf{Precision} & \textbf{Recall} & \textbf{F1-score} \\
\midrule
5  & -- & 0.7500 & \textbf{0.8328} & 0.9496 & \textbf{0.8874} \\
10 & --   & 0.7500 & 0.8257 & 0.9460 & 0.8818 \\
20 & --   & 0.7778 & 0.8231 & 0.9622 & 0.8872 \\
\midrule
5  & 1    & \textbf{0.8333} & 0.7718 & \textbf{0.9856} & 0.8657 \\
5  & 3    & \textbf{0.8333} & 0.8056 & 0.9766 & 0.8829 \\
5  & 5    & 0.8056 & 0.8124 & 0.9658 & 0.8825 \\
\bottomrule
\end{tabular}%
% }
\label{tab:ablation}
\end{table*}

\subsection{Computational Evaluation and Time Costs}
\label{sec:eval}

\begin{table*}[h]

  \caption{Average processing time per sequence (in seconds) and KL/JS divergence between real and simulated pose distributions.}
  \centering
  %---------- (a) Processing time ----------
  \begin{subtable}{0.45\linewidth}
    \centering

    \begin{tabular}{lc}
      \toprule
      \textbf{Component} & \textbf{Time (s)} \\
      \midrule
      Pose Estimation+Vis & 0.31 \\
      DTW Alignment       & 0.37 \\
      Random Forest Sum.  & 0.01 \\
      LLM API Call        & 2.30 \\
      \bottomrule
    \end{tabular}
        \caption{Average processing time per sequence (in seconds).}
    \label{tab:processing}
  \end{subtable}
  \hfill
  %---------- (b) KL / JS divergence ----------
  \begin{subtable}{0.45\linewidth}
    \centering

    \begin{tabular}{lcc}
      \toprule
      Axis & KL$\big(\text{real}\,\|\,\text{sim}\big)$ & JS(real, sim) \\
      \midrule
      X         & 0.1128 & 0.0260 \\
      Y         & 0.1709 & 0.0392 \\
      Z         & 0.0963 & 0.0219 \\
      All (XYZ) & 0.0663 & 0.0132 \\
      \bottomrule
    \end{tabular}
        \caption{Divergence between the empirical Euler-angle distributions of real and simulated poses.}
    \label{tab:pose_divergence}
  \end{subtable}

\end{table*}

\textbf{Computational Evaluation.} To assess the effectiveness of our verbal feedback module, we conducted a computational evaluation of the Random Forest classifier. 
We ran an ablation over two hyperparameters, \texttt{n\_estimators} and \texttt{max\_depth}, using our synthesized dataset (Sec.~\ref{sec:verbalfeedback}). 
Table~\ref{tab:ablation} reports accuracy, precision, recall, and F1-score across settings.  The best configuration (\texttt{n\_estimators}=5, \texttt{max\_depth}=\texttt{None}) achieved 
average precision $0.8328$, recall $0.9496$, and F1-score $0.8874$, which we adopt in our system. 
This performance demonstrates high recall to capture as many critical joint differences as possible while maintaining strong precision to avoid overwhelming users with excessive corrections.

\cglee{To assess how realistic these synthetic pairs are relative to real expert motion, we compare the Euler-angle distributions of real and simulated poses, since no public dataset provides paired expert and learner executions for our target skills.
Figure~\ref{fig:simulation} visualizes the per-axis angle distributions for real versus simulated data, and Table~\ref{tab:pose_divergence} reports their Kullback--Leibler (KL) \cite{kullback1951information} and Jensen--Shannon (JS) \cite{lin2002divergence} divergences. 
Both metrics measure how far two probability distributions deviate from each other, with lower values indicating a closer match. In our case, the synthetic poses remain statistically close to the real motion distribution, yet the injected perturbations still introduce differences from the original motions.}

\textbf{Time Costs.} We implemented four sequential steps: Instruction (3D avatar), Modeling (pose matching), Rehearsal (practice/recording), and Feedback (visual, timing, verbal). Processing times on average are 0.31 s (pose estimation + visualization), 0.37 s (DTW), 0.01 s (random forest), and 2.3 s (LLM API).  To support real-time feedback, we use an efficient pose estimation model~\cite{aoyagi2022development} that runs at 30 FPS on iPhone 16, though higher-accuracy models can be integrated if needed.

%% file: Source/6_user_study.tex
\section{USER STUDIES}
We conducted two user studies with basketball players to evaluate verbal feedback accuracy and the usability and engagement of \name. The first study compared AI-generated verbal feedback with a real coach’s feedback after participants viewed both ideal and user videos. 
% The goal of this study is to evaluate whether our coaching LLM agent can give verbal feedback like coaches. 
\cglee{The goal of this study is to assess how participants perceive the quality of our coaching LLM’s verbal feedback compared to a human coach.}
The second study compared traditional practice (e.g., self-observation with video) and our AR system with a virtual coach, evaluating how each condition helped users identify and correct their errors. Both studies also assessed usability, engagement, and overall user experience.

% \begin{table}[h]
% \centering
% \caption{Ablation Study on \texttt{n\_estimators} and \texttt{max\_depth} for Random Forest. ``\textemdash'' indicates that \texttt{max\_depth} is set to None, meaning that the tree depth is unlimited.}
% \begin{tabular}{cccccc}
% \toprule
% \textbf{n\_estimators} & \textbf{max\_depth} & \textbf{Accuracy} & \textbf{Precision} & \textbf{Recall} & \textbf{F1-score} \\
% \midrule
% 5  & -- & 0.7500 & \textbf{0.8328} & 0.9496 & \textbf{0.8874} \\
% 10 & --   & 0.7500 & 0.8257 & 0.9460 & 0.8818 \\
% 20 & --   & 0.7778 & 0.8231 & 0.9622 & 0.8872 \\
% \midrule
% 5  & 1    & \textbf{0.8333} & 0.7718 & \textbf{0.9856} & 0.8657 \\
% 5  & 3    & \textbf{0.8333} & 0.8056 & 0.9766 & 0.8829 \\
% 5  & 5    & 0.8056 & 0.8124 & 0.9658 & 0.8825 \\
% \bottomrule
% \end{tabular}
% \label{tab:ablation}
% \end{table}

\subsection{Participants \& Experiment Set-up}
\cglee{We recruited 16 basketball players (P1–P16; M = 16, Age: 20–34) via university mailing lists. Participants averaged 6.6 years of experience (range: 2–18) and were categorized as beginner (1), intermediate (12), and advanced (3)}. All had experience practicing alone or with peers. Since \namespace targets all levels, we intentionally recruited participants from a broad skill range to reflect diverse user needs. The study was conducted on an indoor court and lasted 60–75 minutes per participant. Each received a \$20 gift card. We selected five basketball skill videos from online coaching videos \cite{cross_natch, Spin_seal, Cross_over} and recorded corresponding performances to create five video pairs. These were used to obtain both expert and AI-generated feedback. Two actions (e.g., spin seal and cross over) were also used in the second study as the practice tasks.

% CrossSnatch, TimHardaway: https://www.youtube.com/shorts/kUumL0ObMwg
% SpinSimmy(Spinseal), Shamgod:https://www.youtube.com/shorts/2zWWVyFjmew
% Crossover:https://www.youtube.com/watch?v=CGRJdBA6hGc

% Youtube
% Setting

\subsection{Study Procedure}
\textbf{Introduction \& Pre-survey (10 mins).} We provided the participants with an overview of the study and obtained the consent form, and collected background information (level of basketball skill and years of playing experience).

\textbf{Task 1: Verbal Feedback Evaluation (15 mins).} 
The first experiment compared AI-generated and real coach feedback. For AI-generated feedback, we included a baseline (without random forest) and our proposed method (with random forest). Each video pair was paired with three feedback versions: real coach, baseline AI, and proposed AI in randomized order. To reduce bias, participants first completed a verbal quality assessment based on three criteria from our formative study: \textit{clarity}, \textit{identifiability}, and \textit{actionability}. Questionnaires are in Appendix~\ref{questionnaires}. \cglee{To ensure fairness and eliminate ordering bias, three verbal feedback types, (1) real coach, (2) baseline (DTW + LLM), and (3) our proposed Random Forest + LLM, were shown in randomized order per video, with participants unaware of their source.  }

\textbf{Task 2: Skill Training (35 mins).}
To mitigate novelty effects, we intentionally placed the verbal feedback evaluation (Task 1) before the skill training task. 
In the second experiment, participants practiced two basketball actions under two conditions: baseline (self-observation) and \namespace (AR-based virtual coaching). 
A within-subjects design with Latin square randomization was used to balance order and task difficulty. Both conditions were presented in the same AR headset. 
In the baseline, participants watched an ideal performance, practiced twice, and reviewed their recordings to self-correct. In the \namespace, they received step-by-step training and multi-modal feedback from a virtual coach.

\textbf{Post-survey (10 mins).} After completing the two studies, participants were asked to rate the usability, engagement, and overall user experience of \namespace based on key factors \cite{cheng2024viscourt, lee2024sportify, lin2022quest}. In addition, they were asked to compare the two conditions in terms of how effectively they could identify and correct their mistakes. Participants also provided individual evaluations of each feature used in the system.

\subsection{Results}
\subsubsection{Verbal Feedback Preference Results}
As shown in~\autoref{tab:verbal_ranking}, our method received the highest number of first-place rankings (33 out of 80, 41.2\%) and the fewest third-place rankings (7 out of 80, 8.8\%), indicating a consistent overall preference. In contrast, the real coach feedback and baseline were selected first only 28.8\% and 30.0\% of the time, respectively, and received substantially higher third-place rankings (40.0\% and 51.2\%). These results suggest that our model-generated feedback was perceived as more understandable, identifiable, and actionable—criteria directly derived from our formative study. 
\cglee{RF feedback received the highest number of first-place rankings, but a closer look at individual skills shows that preferences varied by motion, with some actions favoring baseline or coach feedback instead (Table~\ref{tab:motion_percent}).}
% This granular feedback demonstrates that our method's effectiveness generalizes well across various skill types, and participants consistently favored RF-generated feedback for its clarity and actionability in ~\autoref{tab:motion_percent}.

\begin{table}[h]
\centering
\caption{Number and percentage of 1st, 2nd, and 3rd place votes for each verbal feedback type.}
\label{tab:verbal_ranking}
\begin{tabular}{lccc}
\toprule
\textbf{Feedback Type} & \textbf{1st Place} & \textbf{2nd Place} & \textbf{3rd Place} \\
\midrule
Real Coach    & 23 (28.8\%) & 16 (20.0\%) & 32 (40.0\%) \\
Baseline      & 24 (30.0\%) & 24 (30.0\%) & 41 (51.2\%) \\
RF (Ours)     & \textbf{33 (41.2\%)} & \textbf{40 (50.0\%)} & \textbf{7 (8.8\%)} \\
\bottomrule
\end{tabular}
\end{table}

\begin{table}[h]
\centering
\caption{Percentage of 1st-place rankings for each feedback type per motion.}
\label{tab:motion_percent}
\begin{tabular}{lccc}
\toprule
\textbf{Motion} & \textbf{Real Coach} & \textbf{Baseline} & \textbf{RF (Ours)} \\
\midrule
CrossOver               & 18.8\% & 18.8\% & \textbf{62.4\%} \\
CrossSnatch             & 37.5\% & 18.8\% & \textbf{43.7\%} \\
ShammgodCisssors        & 18.8\% & \textbf{43.7\%} & 37.5\% \\
SpinSeal                & 31.3\% & \textbf{37.4\%} & 31.3\% \\
TimHardaway In-And-Out  & \textbf{37.4\%} & 31.4\% & 31.4\% \\
\bottomrule
\end{tabular}
\end{table}

% === Ranking Summary for Each Group ===
% Baseline   → First: 13 times, Second: 13 times, Third: 19 times
% Real       → First: 14 times, Second: 10 times, Third: 21 times
% Rf         → First: 18 times, Second: 22 times, Third: 5 times

% === Ranking Summary for Each Motion ===
% CorssOver Baseline   → First: 2 times, Second: 3 times, Third: 4 times
% CorssOver Real       → First: 2 times, Second: 2 times, Third: 5 times
% CorssOver Rf         → First: 5 times, Second: 4 times, Third: 0 times
% CrossSnatch Baseline   → First: 2 times, Second: 3 times, Third: 4 times
% CrossSnatch Real       → First: 3 times, Second: 1 times, Third: 5 times
% CrossSnatch Rf         → First: 4 times, Second: 5 times, Third: 0 times
% ShammgodCisssors Baseline   → First: 4 times, Second: 1 times, Third: 4 times
% ShammgodCisssors Real       → First: 3 times, Second: 4 times, Third: 2 times
% ShammgodCisssors Rf         → First: 2 times, Second: 4 times, Third: 3 times
% SpinSeal Baseline   → First: 3 times, Second: 3 times, Third: 3 times
% SpinSeal Real       → First: 3 times, Second: 1 times, Third: 5 times
% SpinSeal Rf         → First: 3 times, Second: 5 times, Third: 1 times
% TimHardaway_InAndOut Baseline   → First: 2 times, Second: 3 times, Third: 4 times
% TimHardaway_InAndOut Real       → First: 3 times, Second: 2 times, Third: 4 times
% TimHardaway_InAndOut Rf         → First: 4 times, Second: 4 times, Third: 1 times

\subsubsection{Skill Training Performance Comparison: \namespace vs. \ Self-Observation}
We analyzed the accuracy of users' reproduced poses under two conditions: a traditional self-observation baseline and our system, \name. Accuracy was measured as the average angular deviation (in degrees) between the user's joint rotations and the reference motion. Each participant performed two trials per condition. \cglee{In the first trial, participants using \namespace showed lower angular deviation compared to those using self-observation ($M$ = 10.80° vs. $M$ = 14.95°). A Wilcoxon signed-rank test indicated a statistically significant difference between the two conditions ($W$ = 22.0, $p$ = 0.0155). In the second trial, participants again showed lower angular errors with \namespace ($M$ = 11.32°) compared to the self-observation baseline ($M$ = 14.60°). The Wilcoxon signed-rank test similarly revealed a statistically significant difference ($W$ = 21.0, $p$ = 0.0131). }

% Although \namespace showed lower average angular deviation in both trials, the differences did not reach statistical significance.

% np.float64(0.1952), np.float64(0.20300833333333335)] [np.float64(0.24264166666666667), np.float64(0.24349166666666663)
% Shapiro-Wilk Test
% 0.28765189994616547 0.26802027840929554

% Wilcoxon signed-rank test 
% Wilcoxon test statistic: 19.0000, p-value: 0.1294
% Wilcoxon test statistic: 15.0000, p-value: 0.0640
% Ours (degrees): [11.184, 11.632...]
% Self (degrees): [13.899, 13.957]

% Point 1 - Yalong's paper

\begin{figure}[t]
    \centering
    \includegraphics[width=1.0\linewidth]{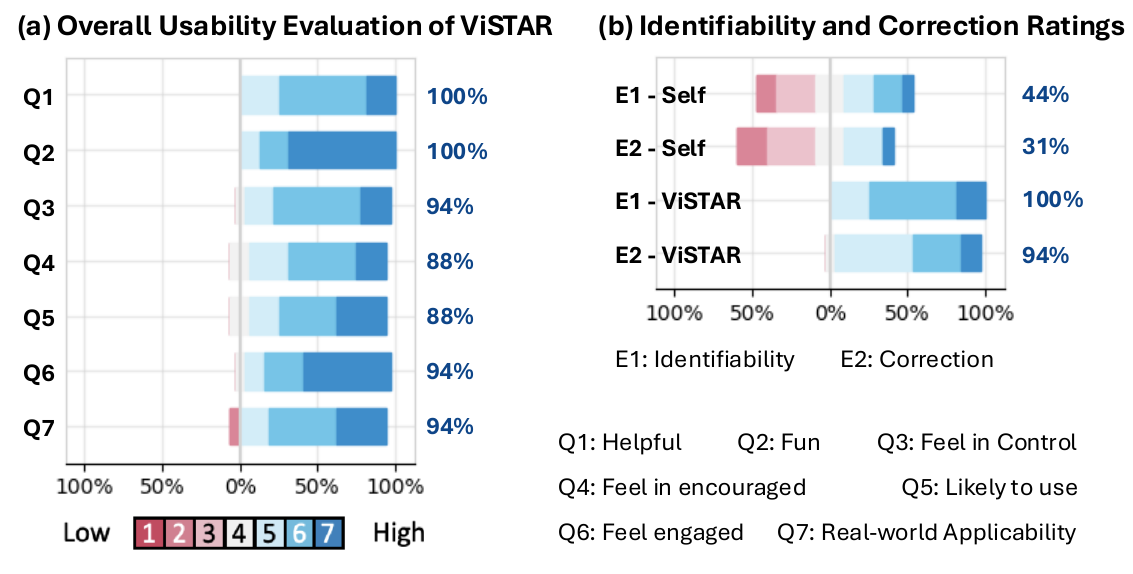}
    \caption{Overall usability ratings of \namespace across seven dimensions (e.g., helpfulness, engagement, applicability), showing high user satisfaction. (b) Comparison of identifiability and correction ratings between the self-observation baseline and \namespace, with \namespace receiving consistently higher scores.}
    \label{fig:result1}
    \Description{Figure 6: Summary of user study results. The figure includes bar charts showing overall usability ratings across multiple dimensions, as well as a comparison of identifiability and correction ratings between a baseline condition and the ViSTAR system.}
\end{figure}

\textbf{\namespace felt more helpful for error recognition and correction.}  
As shown in the right panel of Figure~\ref{fig:result1}, only 44\% and 31\% of participants could identify and correct errors using self-observation. In contrast, 100\% and 94\% of participants reported that \namespace helped them both recognize what went wrong and understand how to fix it. This improvement directly addresses D2 (error recognition) and D3 (receiving actionable feedback). The combination of modeling, visual feedback, and guided practice enabled users to better understand and internalize correct motion.

Participants highlighted that side-by-side avatar comparison in AR made it easier to detect mistakes, as in P3's comment: \textit{“easy to follow the movements because the AR scene closely resembled a real basketball environment.”} Visualizations such as red spheres and 0.25x slow motion helped users inspect details and assess themselves more effectively. For D3 (receiving actionable feedback), verbal feedback increased confidence. P4 noted \textit{“I can trust and correct my motion when my own judgment and verbal feedback are the same.”} Beyond correction, it also prompted reflection. P10 remarked, \textit{“The coaching saying is a new way to think about another aspect while training.”} While one user found some text “too specific,” most described the system as “engaging,” “helpful,” and “great at helping me analyze my movement.”
In additoin, a few participants reported confusion when their own judgment conflicted with the model’s feedback. In such cases, they sometimes hesitated or felt uncertain about which to trust.
Nonetheless, no participants mentioned reconstruction errors as a source of distraction, suggesting that the current accuracy of our pose estimation was sufficient for training purposes. Notably, no participants reported failures of the pose estimation model or raised concerns about system malfunction due to reconstruction errors.

% Weakness different thoughts
% 

\subsubsection{Tool Evaluation of \namespace}
\begin{figure}[t]
    \centering
    \includegraphics[width=1.0\linewidth]{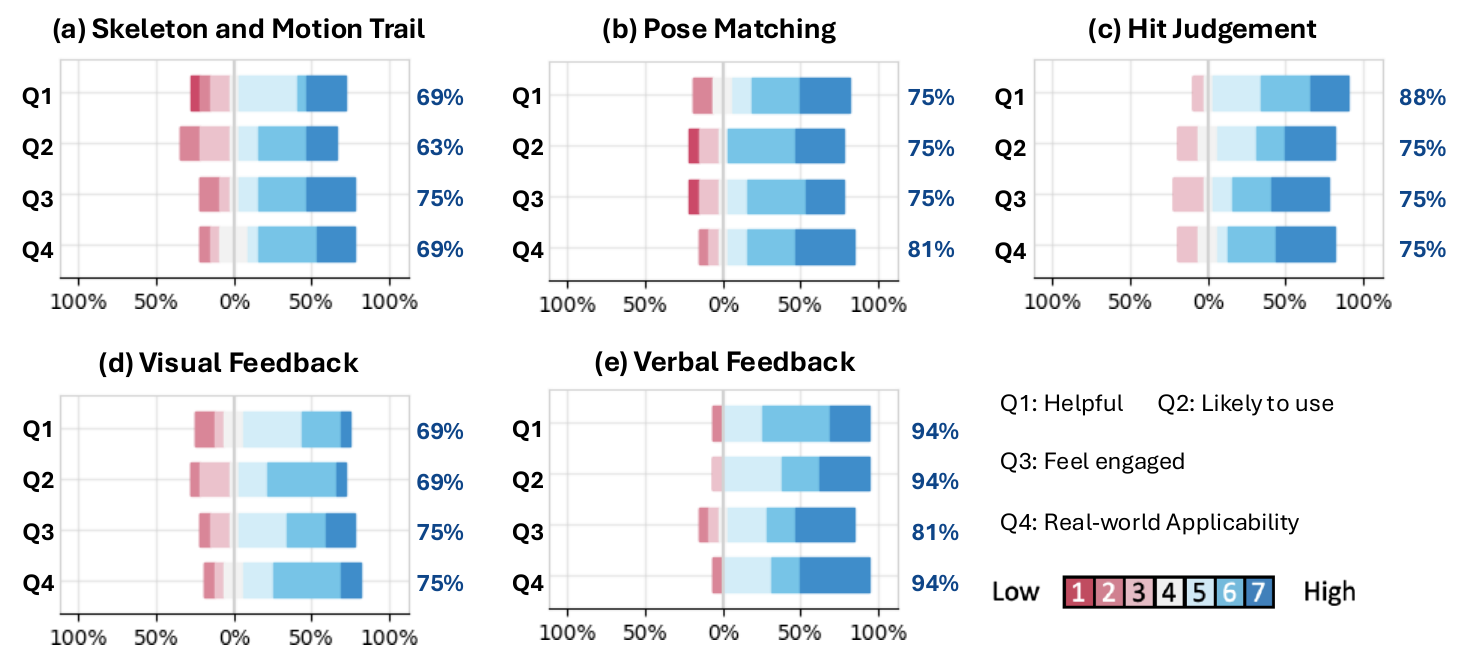}
    \caption{Feature-level evaluation of five feedback components: (a) Skeleton and Motion Trail, (b) Pose Matching, (c) Hit Judgement, (d) Visual Feedback, and (e) Verbal Feedback. User responses reflect consistently high scores across helpfulness, willingness to use, engagement, and applicability.}
    \label{fig:result2}
    \Description{Figure 7: Feature-level evaluation results for different feedback components. The figure compares user ratings for multiple feedback types, including skeletal visualization, pose matching, visual cues, and verbal feedback, across several evaluation dimensions.}
\end{figure}

\autoref{fig:result1} and \autoref{fig:result2} present the results from the second task of our user study, where participants experienced both self-observation (baseline) and \namespace conditions while practicing individual basketball motions. We collected Likert-scale ratings on 1) usability, engagement, 2) identifiability, correction support, and 3) feedback on individual system components.

% \begin{figure}[t]
%     \centering
%     % \includegraphics[width=0.7\linewidth]{images/result1.png}
%     \includegraphics[width=1.0\linewidth]{images/result1.png}
%     \caption{Overall usability ratings of \namespace across seven dimensions (e.g., helpfulness, engagement, applicability), showing high user satisfaction. (b) Comparison of identifiability and correction ratings between the self-observation baseline and \namespace, with \namespace receiving consistently higher scores.}
%     \label{fig:result1}
% \end{figure}

\textbf{\namespace shows high usability and real-world applicability.}  
As shown on the left side of \autoref{fig:result1}, \namespace received high ratings across all usability criteria: helpful (Q1), fun (Q2), feel in control (Q3), encouraged (Q4), likely to use (Q5), feel engaged (Q6), and real-world applicability (Q7). 
% Most items scored in the positive range (5–7) with 100\% agreement, except Q3–Q6 (83–92\%). 
Both Q1 and Q2 received 100\% positive ratings (scores 5–7), while the remaining items also showed positive ratings, ranging from 88\% to 94\%.
These results indicate that \namespace offers an intuitive and engaging experience with promising practical value. 
Open-ended feedback echoed these findings. Participants appreciated the multi-angle views and 3D avatar visualization. For example, P9 said, \textit{``I can see from every angle, this is extremely awesome!''}, while others noted improved perception of spatial relationships and coordination. 
\cglee{These immersive visuals helped them perceive spatial relationships and coordination more clearly than when reviewing flat video.}
% These immersive visuals effectively mitigated the limitations of traditional 2D video learning.

% As shown on the left side of \autoref{fig:result1}, \name received consistently high usability ratings across all seven criteria: helpful (Q1), fun (Q2), feel in control (Q3), encouraged (Q4), likely to use (Q5), feel engaged (Q6), and real-world applicability (Q7). Both Q1 and Q2 received 100\% positive ratings (scores 5–7), while the remaining items also showed strong agreement, ranging from 88\% to 94\%.

\textbf{Each core feature contributes to D1–D3 with 
distinct  strengths and some limitations.}  
\autoref{fig:result2} breaks down user responses across the five features in terms of helpfulness, engagement, and real-world applicability. 
(a) Skeleton and Motion Trail and (b) Pose Matching supported D1 by clarifying complex motions. 
Skeleton and Motion Trail helped users \textit{``see every posture step of movement''} but could be visually dense. P8 found it \textit{``too cluttered''} (scores 63–75\%), suggesting a keyframe or pause-step approach to reduce load. Pose Matching (75–81\%) aided spatial understanding. P2 valued its \textit{“directed feedback”}, though P10 found it \textit{``hard to follow during dynamic motions''}. However, some users noted that these features required effort to interpret during dynamic play, indicating that their utility may be partly tied to the novelty of seeing detailed overlays rather than sustained usability.

% To address D2, (c) Hit Judgement and (d) Visual Feedback provided real-time indication of correctness. Hit Judgement helped users localize precise moments of accuracy or deviation, with P6 sharing that it \textit{``showed real-time where your movement is good and where it is not.''} However, similar to (a), some found its messages difficult to map to specific motion phases. Visual Feedback stood out as the most helpful component overall (69–75\%), with participants describing it as \textit{``the easiest to understand where the problem was''} and \textit{``I could actually compare my movement to the ideal.''} 
To address D2, (c) Hit Judgement and (d) Visual Feedback provided indications of correctness. Hit Judgement helped users pinpoint when their movement aligned or fell out of sync with the expected timing, with P6 noting that it \textit{``showed where your movement was good and where it was not''} in relation to the intended timing. 
Visual Feedback was also reported as helpful (69–75\%), with participants describing it as \textit{``the easiest to understand where the problem was''} and \textit{``I could actually compare my movement to the ideal.''}
Still, when multiple red spheres appeared, some users reported difficulty focusing their attention. Yet a few participants found the cues difficult to map to specific motion phases, indicating that while generally helpful, some aspects require refinement for clarity and pacing.

(e) Verbal Feedback played a particularly strong role in addressing D3 by offering clear and actionable guidance. It received high positive ratings across items. Users praised it as \textit{``The feedback is reasonable and coherent and concise. It's easy to understand.''(P5).}% , and even said it \textit{``felt like there is a live coach'' (P3)}. 
Still, one participant felt there was too much to read and suggested, \textit{``I really liked the verbal feedback! Maybe reduce the text a bit and use bullet points,''} implying that a more concise summary format could enhance usability. Taken together, these modules offered complementary strengths across D1–D3. While users responded positively to all components, common themes around visual clutter and pacing suggest opportunities to enhance clarity and reduce overload.

%% file: Source/7_discussion.tex
\section{DISCUSSION AND FUTURE WORK}
\cglee{We position \namespace\ primarily as an enabling AR+AI platform and discuss the design implications and insights about how such systems can support bodily skill learning and self-guided practice.}

\subsection{Multimodal Feedback as a Key Enabler for AR+AI Coaching}
% \textbf{Beyond Precision: Toward Personalized and Authentic Feedback.} Participants responded positively to the AI-generated feedback, with some saying it \textit{“felt like there is a live coach.”} This supports prior findings that the system improves self-awareness and allows users gain confidence\cite{Weng2025Bridging}. However, one participant (P4) expressed skepticism, stating, \textit{"I'm not sure I fully trust the feedback since it’s AI-generated,"} and felt it resembled a general chatbot. The speaking style also lacked the tone of actual athletes, creating dissonance. P4 noted that real coaches’ feedback tended to be high-level and strategic (e.g., “focus on timing”), whereas AI feedback was more detailed and specific (e.g., “shift your weight to the right foot”). They expressed interest in customizing the feedback style to match that of real coaches or athletes. Interestingly, despite these concerns, the same participant rated the AI feedback more favorably in Task 1, suggesting that while the content quality is strong, perceived authenticity remains a challenge. We suggest a human–AI collaboration approach, where human coaches provide strategic and motivational guidance, and AI systems offer precise, consistent feedback. Such human–AI collaboration can enable more trust, engagement, and precise guidance by leveraging human insight alongside AI-driven detail and consistency~\cite{lobo2024should, oh2018lead}.

\cglee{\textbf{Verbal Feedback as a Design Pattern for Motion-to-Language Interfaces.}}
Although our evaluation centered on basketball skills, the verbal feedback pipeline of \namespace highlights how low-level spatio-temporal joint features can be translated into accessible, actionable coaching through natural language. It illustrates one promising way XR systems can connect motion data and language interfaces.
\cglee{Participants described verbal feedback with 3D avatar as \textit{“felt like there is a live coach,”} reflecting a perceived sense of guidance that echoes prior reports of increased self-awareness in AI-supported practice~\cite{weng2025bridging}. At the same time, concerns about authenticity and trust remained.}
One participant (P4) expressed skepticism, stating, \textit{"I'm not sure I fully trust the feedback since it’s AI-generated,"} and felt it resembled a general chatbot. The speaking style also lacked the tone of actual athletes, creating dissonance. P4 noted that real coaches’ feedback tended to be high-level and strategic (e.g., “focus on timing”), whereas AI feedback was more detailed and specific (e.g., “shift your weight to the right foot”). Such trade-offs suggest a broader design principle: balancing precision and personalization in human–AI collaboration, where human coaches provide strategic guidance while AI systems contribute detailed, consistent feedback~\cite{lobo2024should, oh2018lead, lee2020guicomp}. 
\cgleetwo{
Together, these comments point to a specificity--authenticity tension in motion-to-language coaching: increasing kinematic specificity may improve actionability, yet can reduce perceived coach-likeness when tone and granularity diverge from coaching norms.
These observations suggest that future systems may need to co-tune \emph{what} is said and \emph{how} it is said. 
(1) Highly specific, corrective language can be actionable, yet feel less “coach-like” when it mismatches users’ expectations of coaching tone and level. One practical direction is to provide strategic, coach-like cues by default, with optional drill-down details on demand. 
(2) To reduce “chatbot-like” impressions and support trust, verbal feedback should stay accountable to motion evidence—for instance, by briefly grounding claims in the avatar (e.g., highlighting the referenced body segment) and allowing lightweight verification such as replaying the relevant moment or a simple before/after comparison.
}

\textbf{Multi-modal Feedback: Beyond Multi-faceted Feedback.} To further support a sense of authenticity and user trust, future systems can go beyond multi-faceted feedback (e.g., verbal + visual) by integrating multi-modal input channels such as voice commands, gaze tracking, and contextual performance history. These modalities allow the system to infer the user’s intent, attention, and experience level, and in turn, tailor feedback dynamically. In current AR training system, users must interrupt their physical activity to operate AR controllers,
for example, to view feedback, switch perspectives, or advance to the next segment. These interruptions may break immersion and impose physical and cognitive load. By integrating voice commands, users could interact with the system hands-free (e.g., saying “show me again” or “please recoding”), allowing for a more seamless and natural training experience. Such multimodal interaction could greatly enhance the usability and fluidity of AR-based coaching systems 
\cglee{by reducing the friction we observed when participants had to interrupt movement to operate controllers. 
Evaluating how such modalities affect perceived usability and trust is an important direction for future work. Another opportunity lies in generating feedback using vision or video-language models that directly operate on images or videos of the motion. Exploring these end-to-end motion-to-language generation represents a promising direction for future work. }
\cgleetwo{
This suggests multimodality is not merely a convenience feature, but a mechanism for managing intervention timing under ongoing physical activity.
These observations suggest that multimodal input should protect practice continuity: 
(1) hands-free controls (e.g., voice) can reduce mid-drill interruptions for replays or view changes; and  (2) signals like gaze or performance context may help time interventions (speak, defer, or wait for user pull) to avoid over-interrupting.
}

\subsection{Design Challenges in Dynamic Motion: Balancing Visual Clarity and Feedback Precision}
\textbf{Designing for Feedback in Dynamic Motion.}
Prior work \cite{wu2023ar} demonstrated that 3D trajectory cues are effective for relatively linear movements (e.g., shooting, skiing). However, our findings reveal that in dynamic, high-frequency tasks like dribbling, these cues become cluttered and harder to interpret.
Overall, participants found the trajectory trails helpful for understanding motion flow. However, several users reported difficulty interpreting overlapping visual elements, describing the trails as cluttered or vague during fast-paced sequences. This suggests that the effectiveness of trajectory-based feedback diminish in complex motions unless the visualizations are carefully designed to reduce cognitive load. Our hit judgement feature was valued for real-time feedback on timing and quality, but some users were confused by \textit{“too many cues in rapid succession.”} These issues highlight a common tension in motion feedback: detailed feedback can be useful, but without clear localization or temporal anchoring, it risks overwhelming users. 
Future designs could benefit from adaptive visual abstraction strategies such as keyframe simplification, progressive disclosure, or context-sensitive highlighting to reduce overload while maintaining instructional clarity. 
This suggests two design directions: (1) reduce cue density through adaptive abstraction (e.g., keyframe simplification or progressive disclosure), and (2) improve interpretability by anchoring feedback to moments users can localize (e.g., event-/rhythm-based segmentation with context-sensitive highlighting).
\cglee{Although we currently divide expert motion into equal-length segments for hit judgment, consistent with common time-normalization practices in biomechanics  analysis \cite{dicesare2020high, zhu2023development}, this simplification does not explicitly follow the natural rhythm or event structure of the skill. Future work should investigate event- or rhythm-based segmentation schemes that better align feedback with how athletes perceive phases of movement and balance changes.}

\textbf{Balancing Feedback in Dynamic Motion Across Expertise Levels.}
\cglee{Our user study showed that the effectiveness of motion feedback is not uniform across expertise levels. Less experienced players often asked for simpler, more stepwise cues. For example, one 5-year player (P13) felt that the verbal feedback was \textit{“too detailed"} and requested \textit{"one-by-one suggestions,”} while another (P8) proposed that \textit{“high-level comments or bullet points”} would make it easier to act on. In contrast, more experienced players with 10–18 years of basketball experience appreciated richer, more diagnostic guidance: one participant (P5, 18 years) noted that the verbal feedback provided \textit{“detailed comments which are reasonable and coherent,”} and another (P10) emphasized that it helped them \textit{“easily understand exactly what was going wrong.”} These qualitative patterns echo prior work showing that novices benefit more from explicit, directive feedback, whereas advanced learners benefit from more facilitative, reflective guidance and can better handle detailed information~\cite{shute2008focus, skulmowski2022understanding}.} 
From a design perspective, these observations raise questions about \textit{scaffolded and responsive feedback strategies} that evolve with the learner’s trajectory. A natural direction for future work is to calibrate the granularity and framing of feedback as users progress, shifting from directive to more reflective guidance and examining how this affects learning outcomes. Such strategies could also be informed by user states such as cognitive load, confidence, or engagement, enabling systems to serve as interactive learning companions rather than static feedback channels, as suggested by prior work on action execution and observational learning~\cite{lee2024sportify, zhu2023iball}. 
A practical implication is to scaffold feedback over time: (1) default to concise, stepwise cues for novices with optional drill-down, and (2) provide richer diagnostics for advanced users while monitoring overload (e.g., via interaction signals or self-reported difficulty).

% \textbf{Balancing Feedback Across Expertise Levels.} 
% Our findings revealed that the effectiveness of motion feedback is not uniform: novices in our study reported that explicit, directive cues reduced uncertainty (“tell me exactly what to fix”), whereas more experienced players asked for richer detail to refine subtle aspects of their technique. Although we did not systematically manipulate expertise level or design adaptive policies, these contrasting preferences suggest a tension between overload and oversimplification.

% From a design perspective, we see this as an opportunity for \textit{scaffolded and responsive feedback strategies} that evolve with the learner’s trajectory. Rather than treating expertise as a static category, future systems could experiment with calibrating the granularity and framing of feedback as users progress, for example shifting from directive cues toward more reflective guidance. Building on prior work on motor learning and coaching that emphasizes appropriate task difficulty and feedback framing for different skill levels, future studies could explicitly probe how such strategies impact cognitive load, confidence, and engagement, moving toward systems that act less like static feedback channels and more like interactive learning companions.

\begin{figure}[t]
    \centering
    \includegraphics[width=1.0\linewidth]{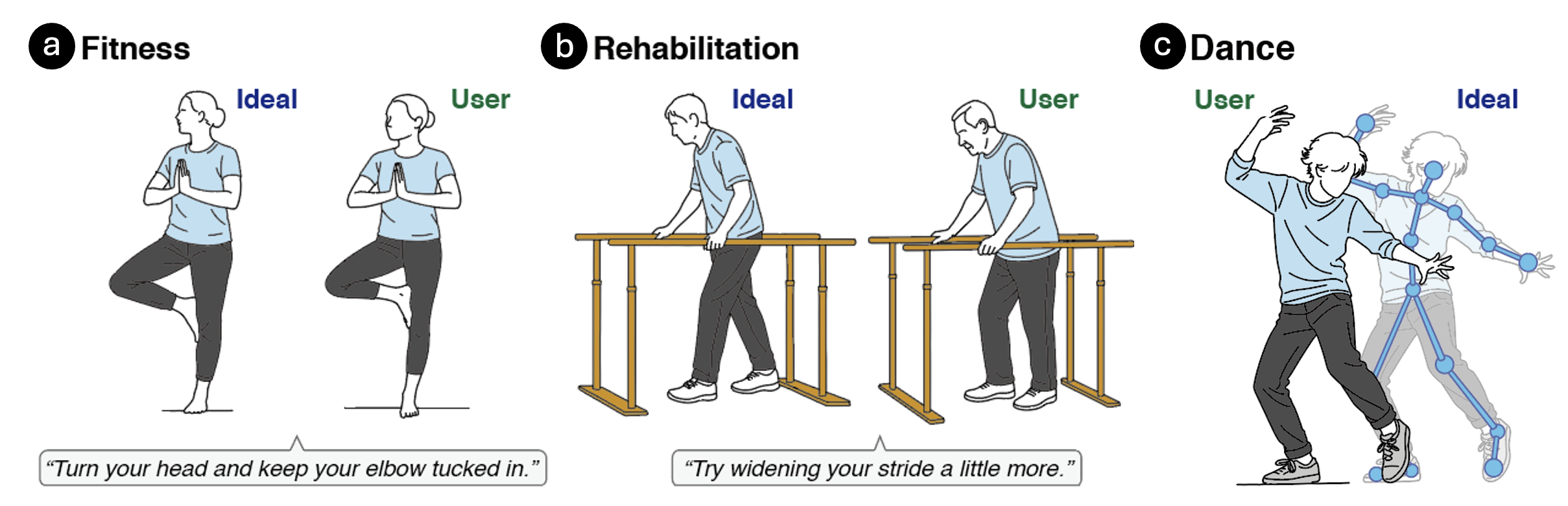}
    \caption{Beyond sports, our system could be applied to diverse domains by comparing user movements against ideal references and providing targeted feedback: fitness (a), rehabilitation (b), and dance (c).}
    \label{fig:domain}
    \Description{Figure 8: Example application scenarios beyond sports. The figure illustrates how the system can be applied to different domains, including fitness, rehabilitation, and dance, by comparing user movements to reference motions and providing feedback.}
\end{figure}

\subsection{Situating Motion Learning: Viewpoint, Social Context, and Beyond Basketball}
\textbf{Viewpoint Alignment and Spatial Understanding in Motion Learning.} Traditional 2D video demonstrations are mirrored, requiring users to mentally reverse left and right when imitating movements. This can cause spatial confusion, particularly in lateral or asymmetric tasks. Prior studies\cite{inoue2021virtual, wu2023ar, diller2025skillar} have suggested that such mirrored views hinder motor learning by increasing cognitive load. In designing our system, we deliberated between a mirrored (coach-facing) or egocentric (user-aligned) avatar view. We adopted the egocentric approach to minimize mental transformation, and participants reported no confusion with left-right alignment. This insight offers valuable implications for future instructional systems in sports, dance, rehabilitation, and other embodied domains. Understanding how different viewpoints (mirrored vs. aligned) shape perception and learning could inform the design of more effective training environments, especially for spatially complex skills. We encourage future studies to empirically compare these visualizations and quantify their impact on learning outcomes across different levels of user expertise.

\textbf{Design Implications for Social and Competitive Learning.}
\cglee{Our system focused on solo skill training and did not capture the reactive and interactive nature of real gameplay\cite{cheng2024viscourt, tsai2017train, tsai2020feasibility}. Participants trained in isolation without the presence of defenders or teammates, limiting opportunities to develop decision-making and adaptability under pressure. Real-world performance depends not only on technical execution} but also on how players respond to others’ movements in dynamic, contested spaces. Future systems could move beyond isolated drills by introducing interactive avatars that simulate opponents or teammates, e.g., dribbling past a defender, coordinating with a teammate, or reacting to tactical shifts. Such designs would better support situational awareness, anticipation, and timing, which are central to competitive play. More broadly, embedding reactive agents into training environments may foster collaboration and strategic thinking, bridging the gap between controlled practice and authentic gameplay. These opportunities extend beyond sports to domains such as dance, rehabilitation, and physical education, where spatial coordination and interpersonal responsiveness are equally critical.

\cglee{\textbf{Generalization to Motion Learning in Other Domains.}
Our current evaluations, datasets, and thresholds are tailored to basketball-specific, isolated motions. However, the same motion-to-language pipeline (i.e., extracting low-level joint data, mapping them to joint descriptors, and generating natural-language feedback) may generalize to other multi-limb coordination tasks in domains such as fitness, rehabilitation, or dance~\cite{ihara2025video2mr}, as conceptually illustrated in \autoref{fig:domain}. Extending the approach would require re-calibrating biomechanical tolerances, redefining task goals in collaboration with domain experts~\cite{lin2023ball}, and validating the system with new participant populations.
}

\subsection{Limitations}
We discuss key limitations of our current implementation and study, \cglee{many of which relate to ecological validity.}

\textbf{Tracking Setup and Alternative Sensing Approaches.}
Our study used a single player in a front-facing capture configuration with one RGB camera and an AR HMD. 
This setup aligns with our target use case of on-court individual skill training before games, where players typically perform short, isolated technical drills alone or with a coach rather than full, dynamic gameplay.
Similar to prior AR/VR sports training work~\cite{gong2024volleynaut, cheng2024viscourt, jheng2025badminton, chen2022vcoach, lin2021towards, esaki2024efficient}, it enabled controlled evaluation of isolated skills. 
However, it does not capture multi-player or crowded scenarios, where player--player and ball occlusions would be more frequent and the HMD’s limited field of view would further constrain ecological realism. 
In addition, our current prototype lacks a dedicated ball-tracking module and does not model defenders or teammates, which prevents full alignment between ball trajectory and the 3D avatar, limits analysis of ball--body relations (e.g., release timing, catch alignment) and multi-player scenarios, and restricts claims about decision-making under pressure.

\cgleetwo{
An efficient alternative to our external video-based tracking is to estimate body pose directly from AR headsets using on-device sensing (e.g., inside-out cameras and IMUs), which could reduce setup burden and improve portability for on-court drills.
However, current headset-only approaches may provide limited full-body fidelity under fast sports motions and frequent occlusions.
In the present work, We chose a single external RGB camera (e.g., a smartphone) to keep the setup lightweight and accessible using readily available, off-the-shelf equipment, without specialized sensors or additional instrumentation.
}

\cglee{\textbf{Study Duration and Headset Comfort.}}
Each participant completed the training session within a single day, which constrained the amount of time they could spend practicing, adapting to the system, and incorporating the feedback. 
\cglee{Although the system provides a highly immersive experience, the current HMD is relatively heavy. 
Therefore, we did not study longer, sweat-heavy, high-speed basketball sessions, where comfort, robustness, safety, occlusion, field-of-view, and ball–interaction issues with head-mounted displays may become more pronounced.}

\cglee{
With 16 participants, our results should be interpreted as providing exploratory trends rather than strong statistical claims. Future longitudinal studies with larger samples, repeated sessions on-court deployment in more realistic training scenarios,
and lighter AR form factors (e.g., glasses-style devices)
are needed to better capture how the system supports learning progression and sustained performance improvement.
}
\cgleetwo{
Our findings should be interpreted as exploratory evidence from a short, single-session study in a constrained setting, rather than as proof of training gains in real-world basketball practice.
}

\cglee{
\textbf{Simplified Feedback Features and Perturbation-Based Modeling.}
In our current design, we de-emphasize spine and torso joints in verbal feedback, prioritizing limb coordination (e.g., foot placement, hand timing), which participants found harder to notice without explicit cues.
Similar metric choices appear in prior systems that focus error descriptors on limb joints~\cite{elsayed2022understanding,debarba2018augmented}. However, this simplification also limits how fully the system can address whole-body coordination and balance, where torso angle plays a critical role. Incorporating richer torso descriptors and linking them to bodily sensations is an important direction for future work. 
Our Random Forest classifier is trained on synthetic motion pairs generated by injecting joint-level perturbations into expert motions. 
As quantified in Sec.~\ref{sec:eval}, these simulated poses are close to the real expert motion distribution, but they still provide only a first approximation to how real learners move and make mistakes. 
Incorporating collecting paired expert–learner motion data that capture realistic error patterns are important directions for future work.
}

%% file: Source/8_conclusion.tex
\section{CONCLUSION}

We presented \name, an AR training system that integrates 3D motion reconstruction with LLM-based verbal feedback to support self-guided athletic practice. \name uses the Behavioral Skills Training (BST) framework to organize visual and verbal feedback. In our study, \namespace yielded lower angular errors than traditional self-observation, but we treat these performance results as exploratory. Participants generally preferred AI-generated feedback over real coaches’, citing its clarity and actionability, and rated \namespace as usable and engaging. Verbal and visual feedback were especially helpful for noticing and interpreting errors. Overall, our results highlight the promise of multi-faceted feedback for motor skill learning and suggest design considerations such as egocentric viewpoints and adapting feedback to user expertise. We position \name as an enabling AR+AI platform that supports reflection on embodied aspects of skill learning rather than a fully validated training intervention.

%% file: Source/9_appendix.tex
\clearpage
\onecolumn
\appendix
\section{Appendix}

\subsection{Task 1: Questionnaires for Verbal Feedback Evaluation}
\label{questionnaires}
\subsubsection{Real Coach's Feedback}

\begin{itemize}
  \item Look up! Hold the hang a bit longer before going into the leg-through. Open up your upper body more.
  \item Look up! Focusing on vertical movement of your center of gravity will improve your motion. After the leg-through, imagine pushing off with your right foot and pulling through with your left foot. (Cross Drive!)
  \item Look up from a lowered head position. Try shifting from a smooth motion to a more rhythmic movement. Keep your upper body more upright while moving.
  \item Try to look up with a wider base. Focusing on vertical movement of your center of gravity will improve your motion.
  \item Look up. Keep your upper body more upright while moving. Try adding an eye fake when doing the in-and-out move.
\end{itemize} 

\subsubsection{Verbal Feedback from Baseline}
\begin{itemize}
  \item You're moving your left elbow and knee too much to the side. Also, there's excessive movement in the right knee and ankle. Try to limit the sideways movement of your left arm and focus on more controlled, straight movements with your right leg for better balance.
  \item Your leg movements, particularly with your knees and ankles, are too excessive. It's particularly notable in the right movements of your right knee and left movements of your right ankle. For the upper body, there are frequent excessive movements with your left shoulder, specifically towards the left and downwards. Your left elbow too shows similar trends. You can improve by limiting these movements - try to be more controlled and precise, especially when moving to the right with your knee and to the left with your shoulder and elbow.
  \item Your leg movements seem to be too much and in the wrong direction. This might be causing bad balance and coordination. Try focusing on moving your knees a bit more to the right and stop moving downwards so much. This change should also help your ankles move properly. On the upper body, your right shoulder and elbow are also moving excessively to the left and down. Pay attention to limiting this movement at a faster pace.
  \item You're moving your arms, especially your elbows, in too many directions. Additionally, your legs, particularly the knees, seem to shift excessively. Try focusing on limiting sideways movement of your elbows and up-down motion for your knees to maintain better balance and coordination.
  \item Your movements are quite excessive in both your legs and arms. For your legs, specifically the knees and ankles, try to limit your movement to the sides and in the up and down directions. In terms of your arms, focus on your shoulders, elbows, and wrists. Try to also limit their movement sideways, up and down, and backwards. Less is more in this case - smaller, controlled movements can significantly improve your coordination and balance.
\end{itemize} 

\subsubsection{Verbal Feedback from \name}

\begin{itemize}
  \item It looks like the main issue is with your left leg, particularly your foot. You're moving it too much to the left. Try to balance your movement by limiting the motion towards the left. Additionally, ease up your left shoulder movement to the left for a more coordinated movement at a faster pace.
  \item You're overextending your left arm and leaning too far with your upper body. Try to limit the forward movement of your left wrist and the rightward shift of your body. Remember, keep the motion controlled and precise for a better balance.
  \item It seems like you're having issues with moving your left foot and left shoulder too much to the left. Try to limit your motion in that direction for better balance and control. Don't forget, it's important to move smoothly and not overextend any particular movement at a faster pace.
  \item Your left leg is moving excessively upward and to the right, try to reduce these movements. Similarly, moderate the movement of your left arm upwards and to the right to maintain a steady posture.
  \item Your left leg movements are too pronounced - try refining them particularly in the right and downward direction. Similarly, ease movement of your left arm to the right, upwards and backwards for better balance.
\end{itemize}